\documentclass[twocolumn]{aastex63}

\usepackage[T1]{fontenc}
\usepackage{newfloat}
\usepackage[caption=false]{subfig}

\usepackage{float}

\usepackage{graphicx}	
\usepackage{amsmath}	
\usepackage{amssymb}	
\usepackage{amstext}
\usepackage{grffile}

\usepackage{xcolor}

\usepackage{xcolor}

\received{June 1, 2019}
\revised{January 10, 2019}
\accepted{\today}
\submitjournal{ApJ}

\shorttitle{GW Signatures from Compact Object Binaries in the GC}
\shortauthors{Wang et al.}
\graphicspath{{./}{figures/}}
\begin{document}

\title{Gravitational-Wave Signatures from Compact Object Binaries in the Galactic Center\footnote{Released on Oct, 10th, 2020}}

\correspondingauthor{Huiyi Wang}
\email{cherylwang815@g.ucla.edu}

\author[0000-0001-5416-2919]{Huiyi Wang}
\affil{Department of Physics and Astronomy, University of California, Los Angeles, Los Angeles, CA 90095, USA}
\affiliation{Mani L. Bhaumik Institute for Theoretical Physics, University of California, Los Angeles, Los Angeles, CA 90095, USA}

\author[0000-0001-8220-0548]{Alexander P. Stephan}
\affiliation{Department of Physics and Astronomy, University of California, Los Angeles, Los Angeles, CA 90095, USA}
\affiliation{Mani L. Bhaumik Institute for Theoretical Physics, University of California, Los Angeles, Los Angeles, CA 90095, USA}
\affiliation{Department of Astronomy, The Ohio State University, Columbus, OH 43210, USA}
\affiliation{Center for Cosmology and Astroparticle Physics, The Ohio State University, Columbus, OH 43210, USA}

\author[0000-0002-9802-9279]{Smadar Naoz}
\affiliation{Department of Physics and Astronomy, University of California, Los Angeles, Los Angeles, CA 90095, USA}
\affiliation{Mani L. Bhaumik Institute for Theoretical Physics, University of California, Los Angeles, Los Angeles, CA 90095, USA}

\author[0000-0003-0992-0033]{ Bao-Minh Hoang}
\affiliation{Department of Physics and Astronomy, University of California, Los Angeles, Los Angeles, CA 90095, USA}
\affiliation{Mani L. Bhaumik Institute for Theoretical Physics, University of California, Los Angeles, Los Angeles, CA 90095, USA}

\author[0000-0001-5228-6598]{Katelyn Breivik}
\affiliation{Center for Computational Astrophysics, Flatiron Institute, 162 Fifth Ave, New York, NY, 10010, USA}

\begin{abstract}

Almost every galaxy has a supermassive black hole (SMBH) residing at its center, the Milky Way included. Recent studies suggest that these unique places are expected to host a high abundance of stellar and compact object binaries. These binaries form hierarchical triple systems with the SMBH, and  undergo the eccentric Kozai-Lidov (EKL) mechanism. 
Here we estimate the detectability of potential Gravitational Wave emission from these compact objects within the frequency band of the Laser Interferometer Space Antenna (LISA) and Laser Interferometer Gravitational-Wave Observatory (LIGO) and Virgo detectors. 
We generate a post EKL population of stars at the onset of Roche limit crossing and follow their evolution to compact object binaries. As a  proof-of-concept, we adopt two metallicities, solar metallicity ($Z = 0.02$) and $15\%$ of it ($Z = 0.003$). We demonstrate that over the observation timescale of LISA, black hole binaries (BH-BH) and white dwarf binaries provides the most prominent GW sources via the EKL assisted merger channel. Systems involving neutron star are less observable but possibly abundant through different merger channels. Our population synthesis of BH-BH with $Z = 0.02$ ($Z = 0.003$) translate to  $\sim$ $4$ ($24$) events per year with LIGO within a 1 ${\rm Gpc}^3$ sphere. We also estimated the number of binaries visible in the LISA band within the inner parsec of our galactic center (and possibly other galaxies) to be about 14 - 150 WD-WD, 0 - 2 NS-BH, 0.2 - 4 NS-NS, and 0.3 - 20 BH-BH.                            

\end{abstract}

\keywords{Galaxy: center --- binaries: close --- gravitational wave --- stars: black hole, neutron stars, white dwarf}


\section{Introduction} \label{sec:intro}
The recent detection of Gravitational Wave (GW) emission from merging stellar-mass black hole (BH) binaries and neutron star (NS) binaries have expanded our ability to sense the Universe \citep[e.g.,][]{Abbott+16a,Abbott+16b,Abbott+16c,Abbott+16d,Abbott+17a,Abbott+17b,Abbott+17d,Abbott+17c,LIGO+19,Abbott+19c}.  However, it remains challenging to distinguish between the formation channels of these merging sources. Recent studies have emphasized the significant contribution of dynamical formation channels in dense stellar environments to the overall GW signals  \citep[e.g.,][]{Portegies+00,Miller+09,O'Leary+09,Banerjee+10,Downing+11,Antonini+12,Rodriguez+15,Cholis+16,Rodriguez+18,Gondan+18,Lower+18,Randall+18,Hoang+18,Samsing+18,Zevin+19,Hoang+20}. One of the unique places that contributes to this overall GW signals via dynamical formation is the center of galaxies \citep[e.g.,][]{Antonini+12,Petrovich+17,Rodriguez+16a,Rodriguez+16b,Hoang+18,Stephan+19}. 

Almost every galaxy has a supermassive black hole (SMBH) at its center \citep[e.g.,][]{Kormendy+95,Ferrarese+05,Kormendy+13}. The closest SMBH to Earth is  Sagittarius A* with $4 \times 10^6$  $\text{M}_\odot$ at the center of our Milky Way Galaxy \citep[e.g.,][]{Ghez+05,Gillessen+09}. Hence detailed observations of the Galactic Center (GC) can provide  valuable insights into the environment and dynamics that must exist in other galaxies as well. 
 
Surrounding the  SMBH at the center of our galaxy is a  dense environment called the nuclear star cluster, which include populations of mostly old stars ($\gtrsim 1$~Gyrs)  \citep[e.g.,][]{Lu+09,Bartko+10,Do+13a,Do+13b,Feldmeier+15,Nogueras+20,Schodel+20} as well as a subset of young stars  ($4-6$~Myr)
\citep[e.g.,][]{schodel+03,Ghez+05,Ghez+08,Gillessen+09,gillessen:2017aa}. 
Binaries that may exists in this nuclear star cluster undergo frequent interactions with neighboring stars as well as gravitational perturbation from the SMBH \citep[e.g.,][]{Heggie+75,Hills+75,Heggie+93,Rasio+95,Heggie+96,Binney+08s,Hopman+09,Hamers+19,Stephan+16,Hoang+18,Rose+20}. These interactions result in astrophysical phenomena such as hypervelocity stars \citep[e.g.,][]{Hills+98,Yu+03,Ginsburg+07} and stellar and compact binary mergers \citep[e.g.,][]{Antonini+10,Antonini+11,Prodan+15,Stephan+16,Stephan+19,Hoang+18}. 
Of course, all these phenomena require high abundance of binaries in the vicinity of the SMBH, and the survival of such binaries depends on the stellar number density \citep[e.g.,][]{Alexander+14,Rose+20}.
 
Binaries are common in our Galaxy with more than half of KGF stars and more than $70$\% of OBA stars having a stellar companion \citep[e.g.,][]{Raghavan+10a}. Thus, it is be reasonable to assume that binaries are also common at the GC. 
Already, there are three confirmed binaries within $\sim$ $0.2$ pc of the GC observed by spectroscopy. The first case is IRS 16SW, an massive eclipsing binary examined by \citet{Ott+99} and \citet{Martins+06}  with $\sim 50$~$\text{M}_\odot$ for each object. Additionally, \citet{Pfuhl+14} reported two binary systems. One is a long-period binary with mass components  $>30$~$\text{M}_\odot$, and the other is an eclipsing Wolf-Rayet binary of $\sim 20$~$\text{M}_\odot$ and $\sim 10$~$\text{M}_\odot$. 
In particular, \citet{Rafelski+07} proposed that the total mass fraction of massive binaries in the GC is comparable to that of the Galaxy's O-stars and is about $7\%$ of the total massive stellar population. Additionally, \citet{Stephan+16} suggested that around $70\%$ of the initial binary population in the GC are expected to remain from the last star formation episode, which occurred $6\, \rm{Myr}$ in the past.

Moreover, the abundant X-ray sources detected within the GC indicate potential stellar companion feeding accreting BHs  \citep[i.e., X-ray binaries e.g.,][]{Muno+05,Cheng+18,Zhu+18,Hailey+19}.On the other hand, \citet{Muno+06,Muno+09} and \citet{Heinke+08} suggested that these X-ray sources could instead be Cataclysmic Variables. 
Additionally, the recent discovery of gas-like objects, the first of those was G2 \citep{Gillessen+12}, suggested the high potential existence of young binary in the GC \citep[e.g.,][]{Witzel+14,Witzel+17,Ciurlo+20}. Lastly, recent work by \citet{Naoz+18} showed that some puzzling properties of the stellar disks could be explained with the existence of binaries.  

Within the vicinity of an SMBH, a stable binary have a tighter orbital configuration than the orbit of its center of mass around the SMBH. In such a system, gravitational perturbations from the SMBH can induce large eccentricities on the binary orbit, known as the "Eccentric Kozai-Lidov" mechanism \citep[EKL, e.g.,][]{Kozai,Lidov,Naoz16}. However, we note that the dense environment of a nuclear star cluster surrounding the GC also provides a high chance of encounters. This may lead to the overall unbinding of binaries or lead to the capturing and hardening of compact object binaries \citep[e.g.,][]{Heggie+75,Heggie+93,Heggie+96,Binney+08s,Rose+20,Hoang+20}.

Recently, \citet{Stephan+19} investigated the dynamical evolution of binary stars in the vicinity of SMBH subject to the EKL, including tidal interactions, general relativity (GR), as well as single and binary stellar evolution. They showed that while 75\% of stellar binaries in the GC that are interacting with the central SMBH become unbound by the interaction, the remaining 25\% will merge after a few Myrs. Of the merging binaries, $\sim 14.6$\% will become compact objects binaries while the remaining $\sim 85.4\%$ will merge while the stellar components are Main Sequence, Red Giant, or Helium Star phase. The final results of \citet{Stephan+19} show that $1.8$\% of those $\sim 14.6$\% compact objects will form a black hole binary (BH-BH), $0.6$\% will form a black hole - neutron star binary (NS-BH), $1.2$\% will become white dwarf - neutron star binaries (WD - NS), $15.1$\% will become white dwarf binaries (WD-WD), while no neutron star binaries (NS-NS) can form.

We note that natal kicks that compact objects binary received during their supernova (SN) explosions also affect the number of binaries formed and their orbital parameters \citep[e.g.,][]{Kalogera00,Bortolas+17}. In the GC where binaries undergo dynamical hierarchy interactions with the SMBH, natal kicks might eject those compact binaries with a high escape velocity to completely unbind the triple systems \citep[e.g.,][]{Michaely+16,Antonini+16,Parker+17,Bortolas+17,Lu+19}. On the other hand, \citet{Lu+19} recently showed that supernova kicks can more often result in shrinking the separation then expanding the orbit, thus contributing to possible GW events.

Currently, terrestrial GW detectors can only observe merging BH or NS binaries during their final inspiral phase. Nevertheless, detections of  GW emission from binaries still in orbit are significant in revealing the binaries' formation history \citep[e.g.,][]{Breivik+16,Nishizawa+17}. Those sources can best be resolved via the Laser Interferometer Space Antenna (LISA), which is sensitive to mHz frequencies. \citep[e.g.,][]{Folkner+98,Pau+17,Robson+18b,Robson+19}. Observation of those GW sources can potentially contribute to our understanding of close binary evolution, the distribution of X-ray sources, supernova (SN) explosions, Gamma-ray bursts as well as galactic structure \citep[e.g.,][]{Yu+10}. Fortunately, binaries close to the SMBH exhibit measurable eccentricity oscillation due to the EKL cycle, and are thus distinctive from isolated field binaries. \citep[e.g.,][]{Hoang+19,Randall+19,Emami+20,Deme+20}.

The paper is organized as follows. We first provide the basic equations for estimating the signal to noise ratio (\S \ref{sec:2}). Next, we analyze the potential detectability of compact objects within the vicinity of the GC while being agnostic to the formation mechanism (\S \ref{sec:3}). Then we generate a large population of binaries at the onset of their Roche limit crossing following their EKL evolution(\S \ref{sec:4.1}). Next, we evolve these binaries using the {\tt COSMIC} \citep{Katie+19} code into compact object binaries (\S \ref{sec:4.2}) and investigate the resulting GW signatures in terms of LISA sensitivity curve (\S \ref{sec:4.3}). We offer a crude approximation of the the LIGO detection rate in \S \ref{eligo} and a final discussion in \S \ref{sec:dis}.
 
\section{Basic equations for Signal-to-Noise}\label{sec:2}

For completeness we specify the relevant equations for calculating the LISA SNR, see \citet[]{Kocsis+12,Robson+19} and \citet{Hoang+19} for a complete derivation. The SNR of a binary, with a semi-major axis $a$ and eccentricity $e$ is given by 
\begin{equation}\label{eq:SNRS}
< {\rm SNR}^2(a,e)> =  \frac{16}{5}\int \frac{ |\tilde{h}(a,e,f)|^2}{S_n(f)} df \ , 
\end{equation} 
where, $S_n(f)$ is the 
effective noise power spectral density:
\begin{equation}\label{eq:Sn}
S_n(f)=\frac{P_n(f)}{\mathcal{R}(f)} \ ,
\end{equation}
and has a units of Hz$^{-1}$. $\mathcal{R}(f)$ is the dimensionless sky and polarization averaged signal response function of the instrument and $P_n(f)$ is the power spectral density of the detector noise.

The square of the GW strain amplitude in the frequency domain, $|\tilde{h}(a,e,f)|^2$, from Eq.~(\ref{eq:SNRS}), is the sum of the strain amplitude at each orbital frequency harmonic, $ f_n$. In other words,

\begin{equation}\label{eq:hfsq}
|\tilde{h}(a,e,f)|^2 \approx \sum_{n = 1}^{\infty} h^2_n(a,e,n) T_{\rm obs}^2 \Big(\frac{\sin(\pi(f_n - f) T_{\rm obs})}{\pi (f_n - f) T_{\rm obs}}\Big)^2 \ ,
\end{equation}
\citep[e.g.,][]{Kocsis+12,Hoang+19}, where, 
\begin{equation}\label{hn}
    h_n(a,e,f_n) = \frac{2}{n}\sqrt{g(n,e)} h_0(a) \ ,
\end{equation}
\begin{equation}\label{eqn:ho}
    h_0 (a)  =\sqrt{\frac{32}{5}}\frac{G^2}{c^4}\frac{m_1 m_2}{D_l a} \ ,
\end{equation}
and 
\begin{equation}\label{gn}
\begin{split}
        g(n,e) = & \frac{n^4}{32}\Biggl[\left(J_{n-2} - 2eJ_{n-1} + \frac{2}{n}J_n + 2eJ_{n+1} - J_{n+2}\right)^2  \\ &+ (1-e^2)(J_{n-2} - 2J_{n} + J_{n+2})^2 + \frac{4}{3n^2}J_{n}^2 \Biggr],
\end{split}
\end{equation}
where $D_l$ is the luminosity distance between the source and detector, $T_{\rm obs}$ is the observation time of each binary source,  $G$ is the Newton's gravitational constant, and $J_i$ is the $i$th Bessel function evaluated at each $ne$ \citep{Peters+63}.
The frequency harmonic, $f_n$ in Equation (\ref{eq:hfsq}) is defined as $f_n = nf_{\rm orb}$, where 
\begin{equation}\label{f_orb}
    f_{\rm orb}(a) = \frac{1}{2\pi}\sqrt{\frac{G(m_1 + m_2)}{a^3}} \ .
\end{equation}
The full width at half maximum (FWHM) of the integral from Equation (\ref{eq:hfsq}) gives,
$$\int \text{sinc} (\pi (f_n - f) T_{\rm obs})^2 df \approx 0.885895/T_{\rm obs}$$ 
Thus, combining Equation (\ref{eq:SNRS}), (\ref{eq:hfsq}) and (\ref{hn}), the final expression of the SNR of LISA can be approximate as 
\begin{equation}\label{eq:snr}
    < \text{SNR} (a, e) >\approx \frac{8}{\sqrt{5}} h_0(a) \sqrt{ 0.885895 \times T_{\rm obs} \sum_{n_{min}}^{n_{max}} \frac{g(n,e)}{n^2 S_n(f_n)}}  \ .
\end{equation} 

In this paper, we calculate the SNR of each binary system based on the entire LISA mission life time $T_{\rm obs} = 4 $ yr, to accumulate the highest signals. However, for systems which merge within $4\, \rm{yr}$, we take $T_{\rm obs}$ as their merging timescale.

Additionally, the characteristic strain of an evolving binary with eccentric orbits can be crudely approximated with the Fourier Transform of a stationary binary of Equation (\ref{eq:hfsq}) \citep{Hoang+19}:
\begin{equation}\label{strain}
    h_c^2(a, e, f) = 4f^2|\tilde{h}(a,e,f)|^2 \times {\rm min}\left(1, \frac{f_n}{\Dot{f_n}}\frac{1}{\Delta T_{\rm obs}}\right) \ ,
\end{equation}
Here, the factors within the minimum functions are taken due the fact that each signal only accumulated within their own lifetime or the LISA observation timescale \citep[e.g.,][]{Culter+94,Flanagan+98}. If $a$ and $e$ changes insignificantly over the observation timescale of LISA, the signal power only accumulates in each frequency bin. This equation is used as an approximation of the binary's strain curve in the LISA parameter space.

\section{Detectability of sources in the GC via LISA - A Proof-of-concept}\label{sec:3}

While EKL  is one of the processes that can induce compact object binaries, other mechanisms can also drive the formation of compact object binaries. These processes include, but not limited to: captures \citep[e.g.,][]{O'Leary+09,Hoang+20}, hardening by weak interactions \citep[e.g.,][]{Heggie+75,Alexander+14,Rose+20}, 3-body interactions \citep[e.g.,][]{Binney+08s} etc. Thus, as a first step, we offer a proof-of-concept that is agnostic to the binary formation mechanism. 

We investigate six representative examples of different compact objects involving BH, NS, and WD, as listed in Table \ref{comp}. Those examples aim to provide a proof-of-concept of the Signal-to-Noise Ratio (SNR) of the potential compact object binaries within the GC viewing via LISA Detector.

\begin{table}
\caption{Six representative examples of compact object binaries referencing Figure \ref{fig:snr}. The masses of black hole are taken as 30 $\text{M}_\odot$ and 10 $\text{M}_\odot$. The masses of NS and WD are taken to be the typical lower limit of 1.4 $\text{M}_\odot$ and 0.5 $\text{M}_\odot$ respectively.}
\centering
\label{comp}
\begin{tabular}{rcccc}
\noalign{\smallskip} \hline \hline \noalign{\smallskip}
Figure \ref{fig:snr} & $m_1$ ($\text{M}_\odot$) & Type 1 & $m_2$ ($\text{M}_\odot$) & Type 2 \\
\hline
a & 30 & BH & 30 & BH \\
b & 10 & BH & 30 & BH \\
c & 1.4 & NS & 30 & BH \\
d & 1.4 & NS & 10 & BH\\
e & 1.4 & NS & 1.4 & NS \\
f & 0.5 & WD & 0.5 & WD \\
\noalign{\smallskip} \hline \noalign{\smallskip}
\end{tabular}
\end{table}

\begin{figure*}
    \centering
    \subfloat{%
    \includegraphics[width=.4\linewidth]{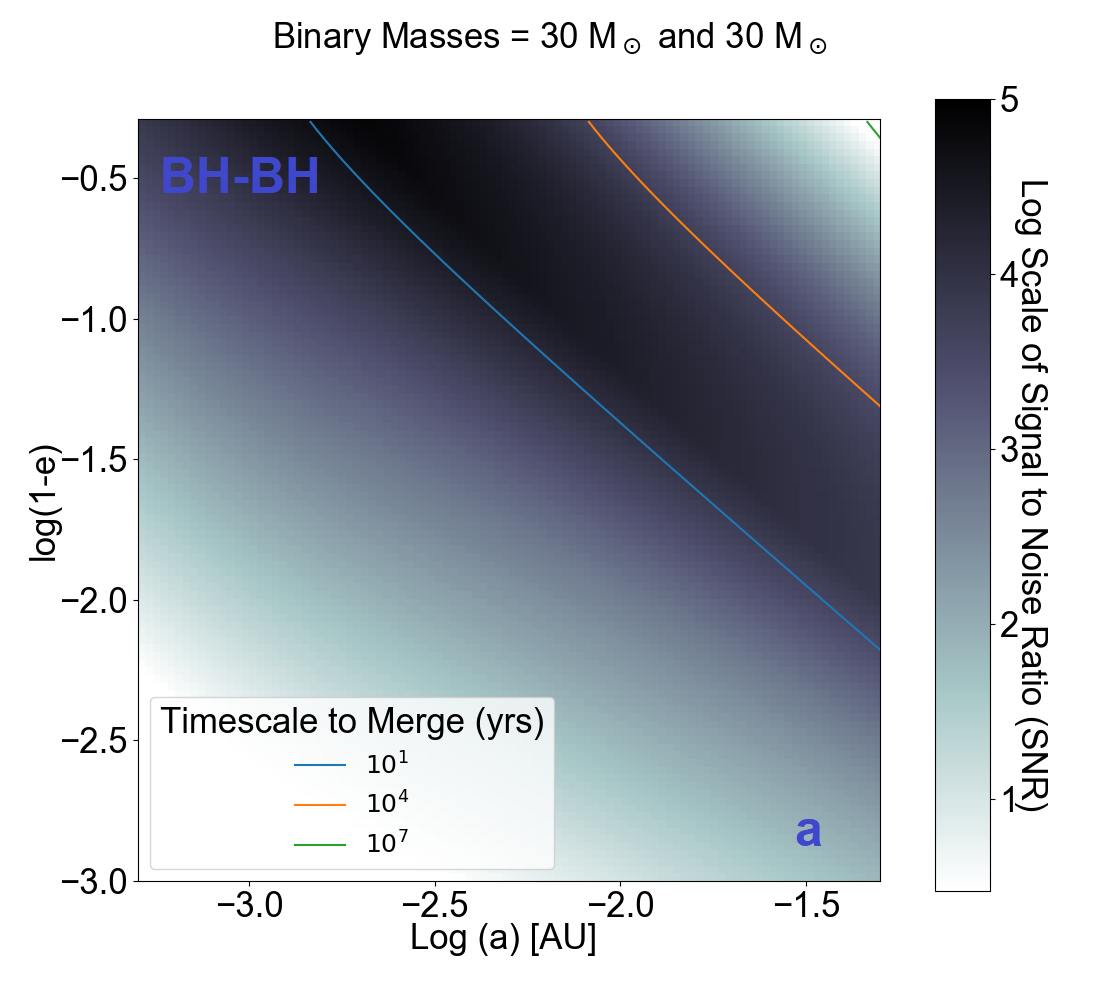}\label{fig:sfig1}
    } \qquad
    \subfloat{
    \includegraphics[width=.4\linewidth]{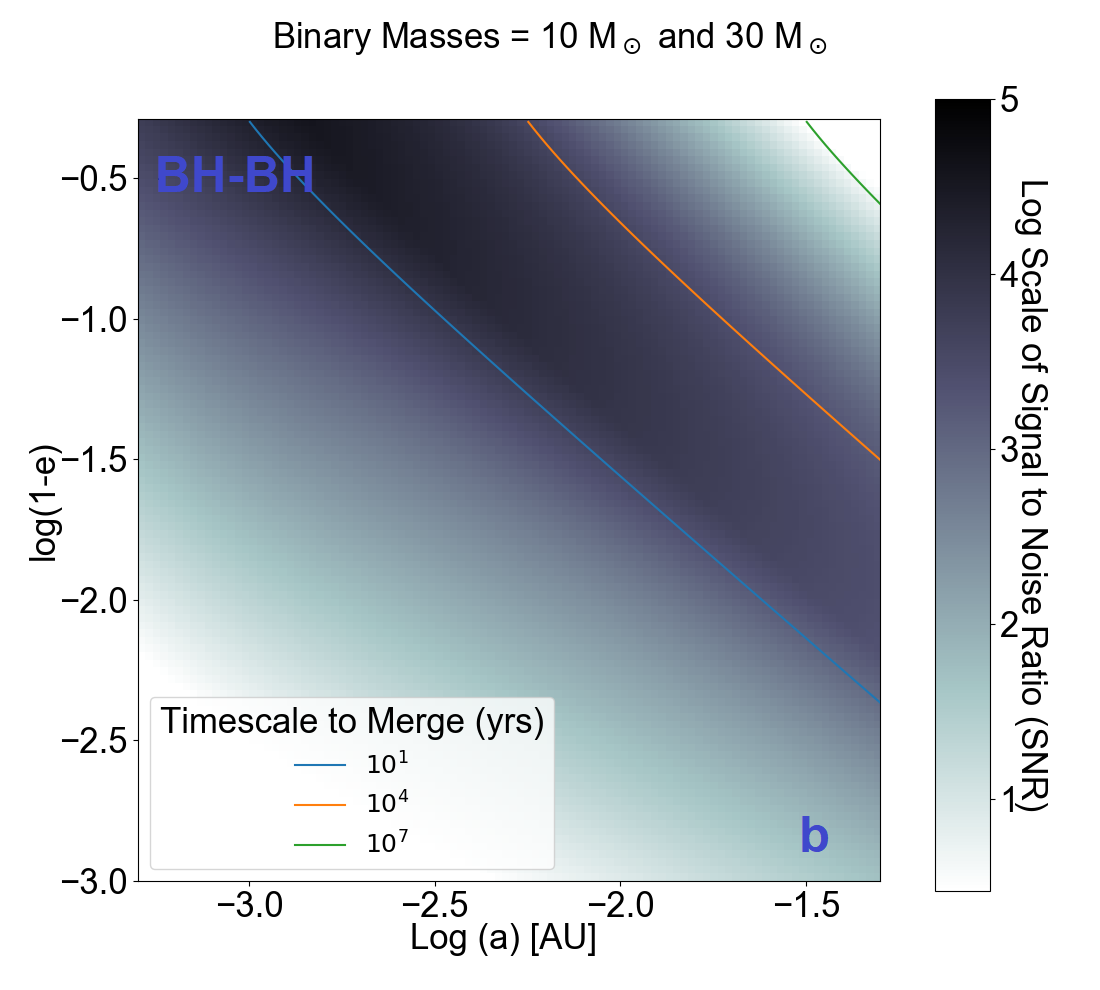}
      \label{fig:sfig2}
    }\\
    \subfloat{
    \includegraphics[width=.4\linewidth]{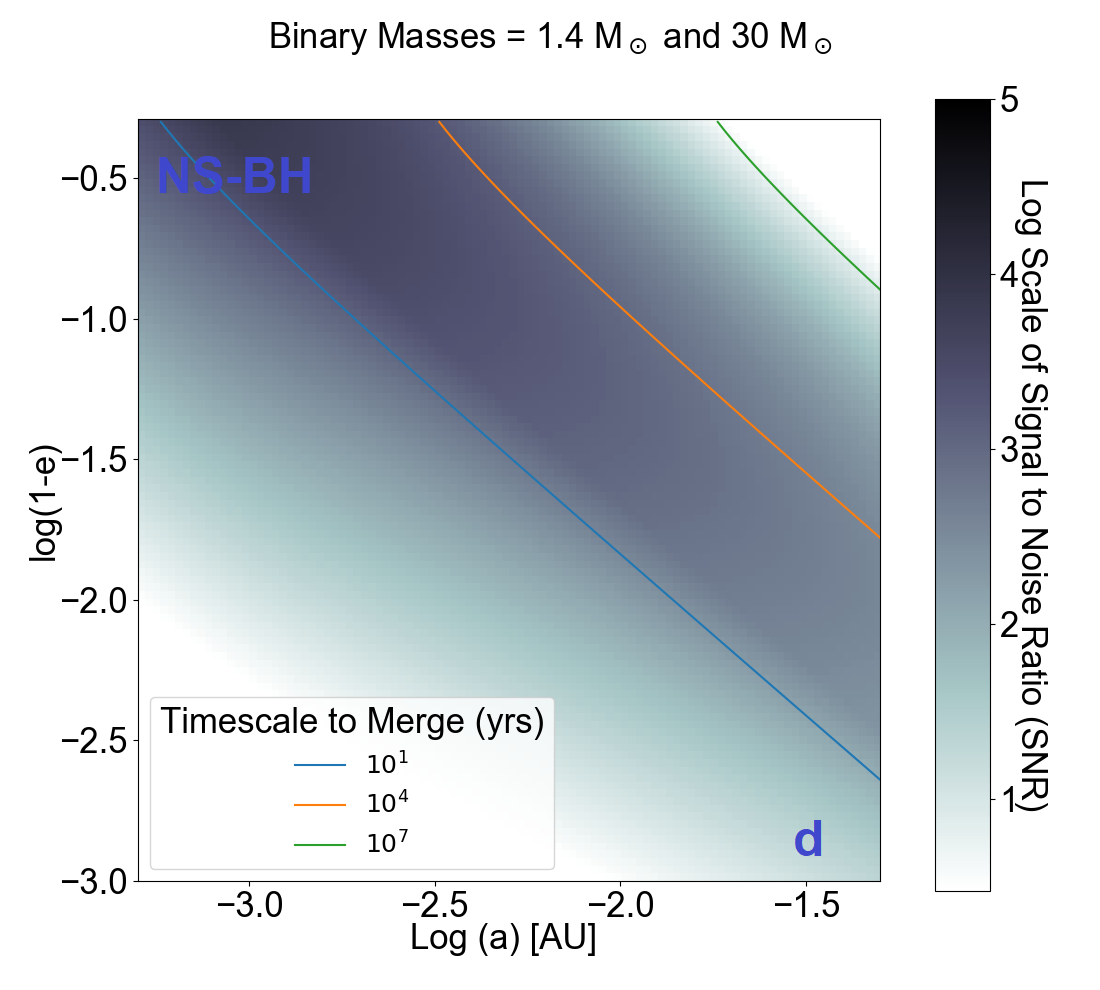}
      \label{fig:sfig3}
    }\qquad
    \subfloat{
    \includegraphics[width=.4\linewidth]{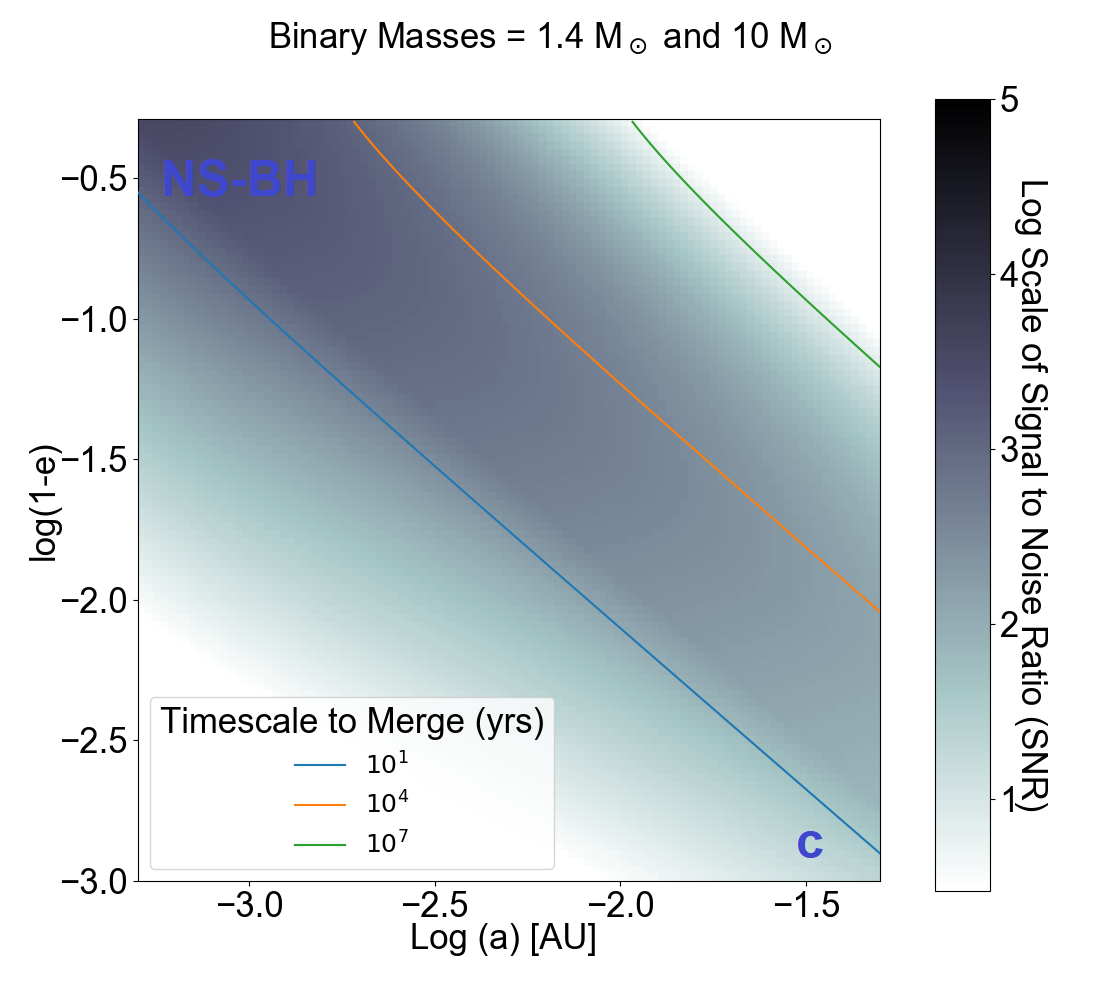}
      \label{fig:sfig4}
    } \\
    \subfloat{
    \includegraphics[width=.4\linewidth]{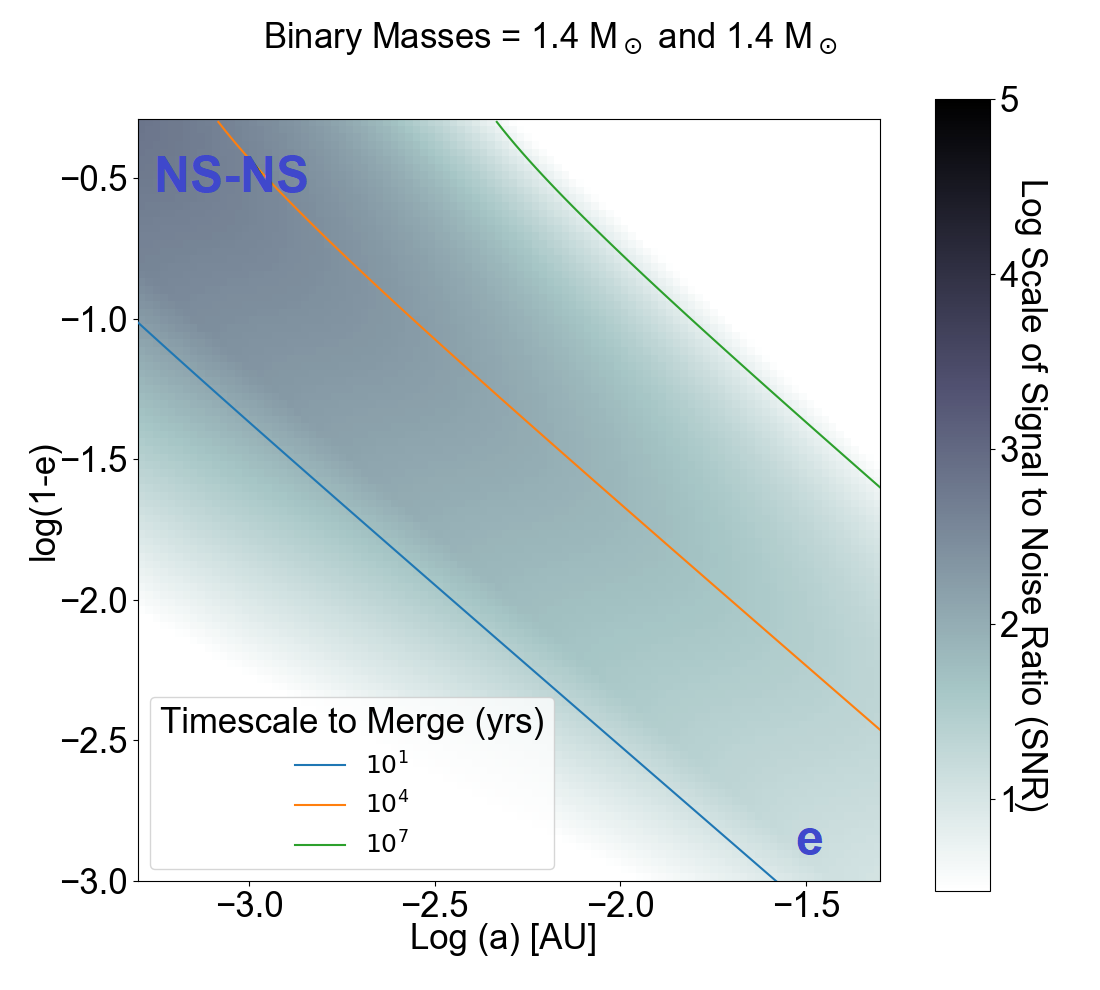}
      \label{fig:sfig5}
    }\qquad
   \subfloat{
   \includegraphics[width=.4\linewidth]{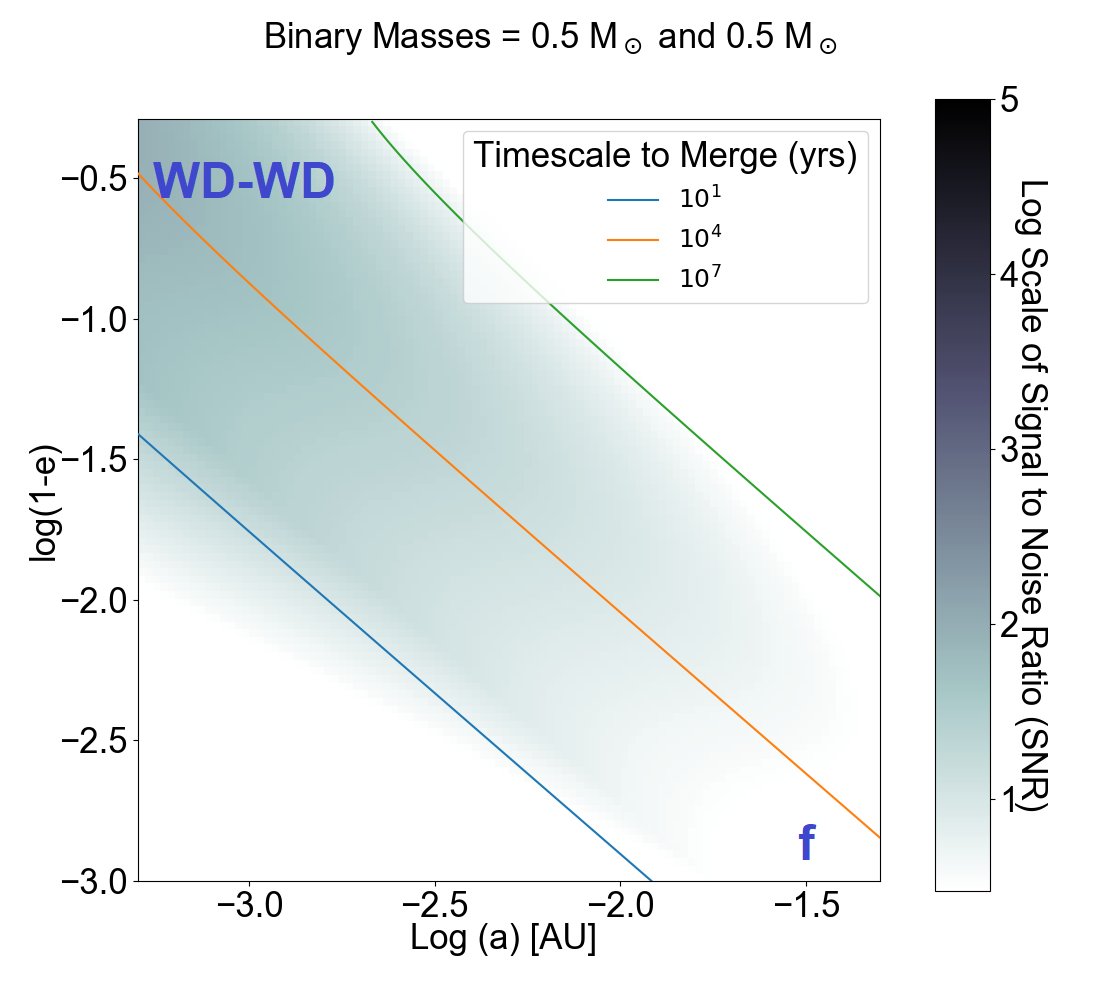}
      \label{fig:sfig6}
    }\\
\caption{We present the Signal-to-Noise Ratio (SNR) density plot of six representative compact object binaries according to Table \ref{comp}. The SNR are plotted with orbital parameter $0.001  < 1 - e < 0.5$ on y-axis and  $0.0005 < a < 0.05$ on x-axis, both with log scale. Each density point is color-coded according to its SNR.  We also overplot three merging timescale in years of $10$, $10^4$, and $10^7$. The luminosity distance ($D_l$) is taken to be $8$ kpc and the observation interval ($T_{\rm obs})$ takes the minimum between each system's merging timescale and LISA mission time ($4$ yr). }
\label{fig:snr}
\end{figure*}

In Figure \ref{fig:snr} we depict the SNR for the six representative examples using Equation (\ref{eq:snr}), considering a wide range of separation and eccentricity\footnote{We expect the eccentricity to always increase, even for hard binaries due to the EKL mechanism \citep{Tey+13,Naoz+14,Li+13,Li+14,Naoz16,Hoang+18,Stephan+16,Stephan+19}.
}. As expected, the SNRs of WD binaries are low ($\lesssim100$, but still larger than $5$) for wide range of the parameter space. The SNRs for NS binaries are mostly above $\sim 100$  while the SNRs for binaries with BH can be as high as $10^5$ with the majority of the parameter space yielding SNRs $>100$.

Those compact object binaries will eventually merge via the timescale estimated as:
\begin{equation}\label{eq:tscale}
\begin{split}
    {t_{\rm GW} }= 1.602 \times 10^5 {\rm yr} \times \left(\frac{{m}_1}{10 {\rm M}_{\odot}}\right)^{-1} \left(\frac{m_2}{10 \text{M}_{\odot}}\right)^{-1} \\ \times \left(\frac{m_1 + m_2}{20 \text{M}_{\odot}}\right)^{-1} \left(\frac{a}{ 0.01\text{AU}} \right)^{4}\times f(e)(1-e^2)^{7/2} \ ,
\end{split}
\end{equation}
\citep[e.g.,][]{Blaes+02}.  In Figure \ref{fig:snr}, we over plot three representative merger times of $10$ yr, $10^4$ yr, $10^7$ yr. 

As represented in Figure \ref{fig:snr}, the BH-BH systems (top row) with the larger SNRs, will merge in less than 10 years, thus appearing as a LIGO signal (not shown here). However, still, a large part of the parameter space, with merger timescales between $10-10^4$~yr, have SNR above $1000$. A strong signal is also depicted in the NS-BH examples (middle row), with SNR above $\sim 500$. The majority of high SNR is expected to have a lifetime larger than 10 years. Although with weaker signals, the NS binaries (bottom left row) still have long-lived systems (merging in $10-10^4$ yr) with relatively large SNR ($\sim 100$). The SNRs of WD binaries (bottom right row) are the lowest and thus span the smallest regions of the parameter space. However, they are still detectable (with SNR of $\sim 50$). 

We note that the signal of a system that merges within $T_{\rm obs} < 4$ yr cannot simply be described as the sum of the harmonics \citep[e.g.][]{Barack+04}, as presented in Figure \ref{fig:snr}. In addition, in these parameter spaces, some large EKL-induced eccentricity oscillations can possibly take place. This situation is explored by \citet{Hoang+19} which showed that EKL eccentricity oscillations can be observed in LISA and may even infer the existence of a SMBH. In Figure \ref{fig:snr}, most of the systems with large SNR lay above the $t_{\rm GW}=10$~yr, which render our use of Equation (\ref{eq:snr}) valid (while the observation time is $4$~yr, we use the $10$~yr line to guide the eye).  

\section{GW signature from an EKL population at the GC}
Here we outline the procedure of population synthesis of binaries at the GC. We begin with adopting binaries' distribution properties from \citet{Stephan+16,Stephan+19} of post-EKL systems at the onset of Roche-limit crossing. We then generate $10^6$ systems at this stage and use {\tt COSMIC} \citep{Katie+19} to evolve them with time. Finally, we calculate the GW signal of the resulting compact object binaries via LISA detector and estimate their LIGO detection rate.

\subsection{Post EKL population }\label{sec:4.1}

As mentioned above, recently \citet{Stephan+16,Stephan+19} investigated the stellar binaries evolution in the vicinity of SMBH including the EKL mechanism, tides, GR, and stellar evolution. For stellar binaries overflowing their Roche limit, the gas flow captured by a stellar companion will results in mass transfer, which renders subsequent stellar evolution different than a single star evolution \citep[e.g.,][]{Hurley+02,Sharpee+12,Toonen+16,Antonini+17}. \citet{Stephan+19} followed this population using the {\tt COSMIC} stellar evolution code (see \S \ref{sec:4.2}) and predicted a high abundance of compact object binaries within the GC. This formation channel seems promising to generate GW sources within GC \citep[e.g.,][]{Antonini+10,Antonini+12,Hoang+18,Stephan+19}.

We employ the same numerical setup and initial distribution as \citet{Stephan+16,Stephan+19} and generate $10^6$ systems at the {\it onset of Roche limit crossing}. Here, we define the Roche limit of a binary system with Equation (\ref{roche}) \citep[taking $m_2 < m_1$ and $\eta = 1.6$, e.g.,][]{Naoz16}, 
\begin{equation}\label{roche}
    R_{\rm oche} = \eta R_2 \left(\frac{m_2}{m_1 + m_2}\right)^{-1/3} \ .
\end{equation}

It is worth describing \citet{Stephan+16,Stephan+19} initial conditions that led to the Roche limit crossing we analyze.  In those studies, the primary stellar initial mass function is taken from \citet{Salpeter+55} with $\alpha = 2.35$ with a mass limit between 1 ${\rm M}_\odot$ and 150 ${\rm M}_\odot$. The mass ratio to the secondary mass in the binary uses that of \citet{Duquennoy+91}. In other words, the mass ratio $m_1/m_2$ was taken from a Gaussian distribution with a mean of $0.23$ and a standard deviation of $0.42$.  The SMBH of Sagittarius A* was set to $4\times 10^6$  ${\rm M}_\odot$ \citep[e.g.,][]{Ghez+05,Gillessen+09}. The inner binary semi-axis distribution uses the same as that of \citet{Duquennoy+91} while the outer orbital period was distributed uniformly in log with a maximum of $0.1$~pc. The eccentricity of the inner binary is uniformly drawn from $0$ to $1$ \citep{Raghavan+10a} and the eccentricity of the outer binary uses a thermal distribution \citep{Jeans+19}. Both the inner and outer argument of periapsides were taken from a uniform distribution between $0$ and $2\pi$. Additionally, those binaries generated must satisfy orbital and analytical stability \citep[See ][for a complete specification]{Naoz+14,Naoz16,Stephan+16,Stephan+19}. Using these initial conditions, \citet{Stephan+16,Stephan+19} evolved $1,500$ binaries via EKL mechanism, tides, GR and single-stellar evolution. When the binaries crossed their Roche-limit, \citet{Stephan+19} used {\tt COSMIC} to follow their binary stellar evolution. 

\begin{figure}
    \centering
    \subfloat{
    \includegraphics[width=0.97\columnwidth]{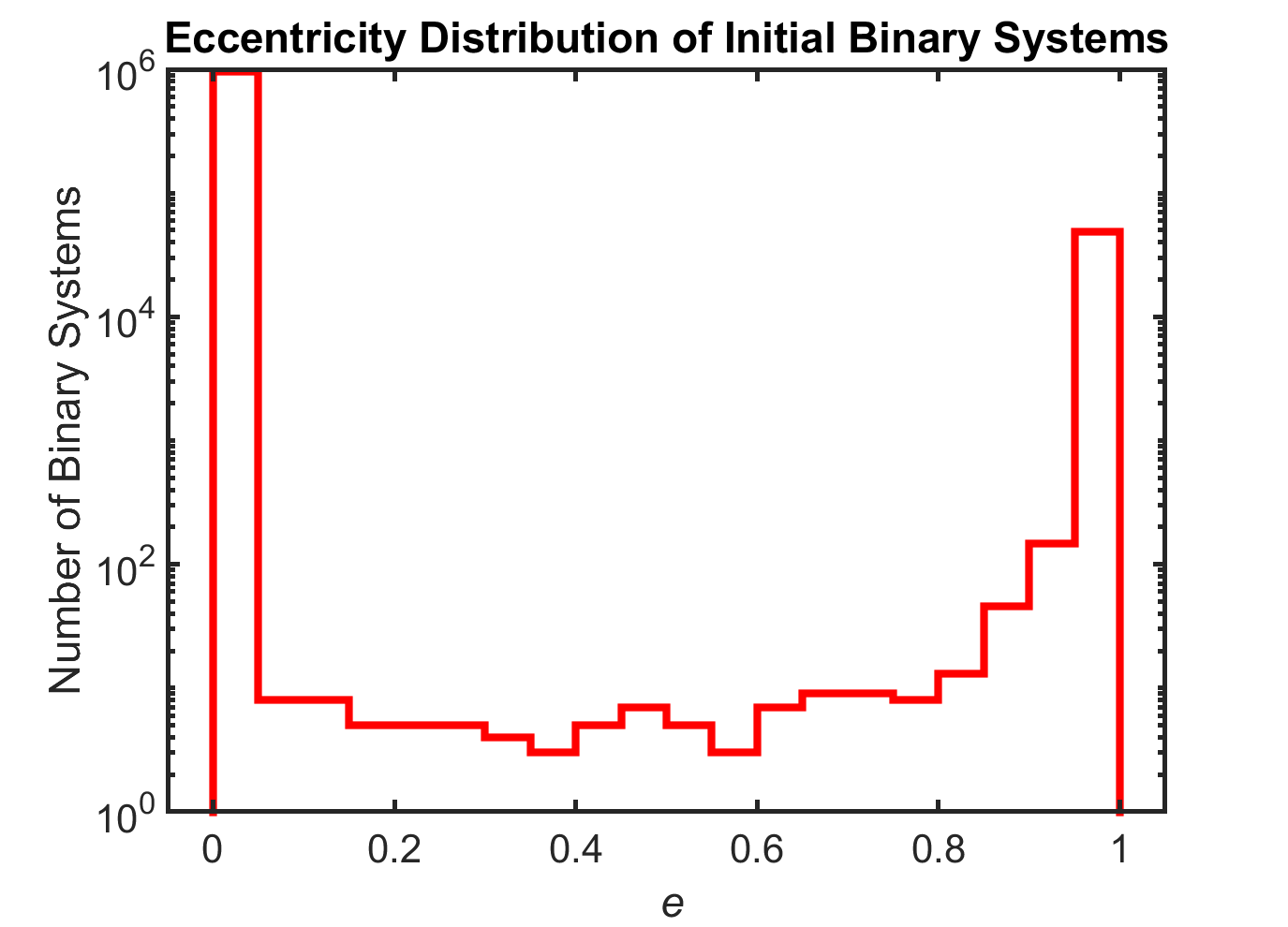}
    \label{fig:inihi}
    }\\
    \subfloat{
      \includegraphics[width =0.97\columnwidth]{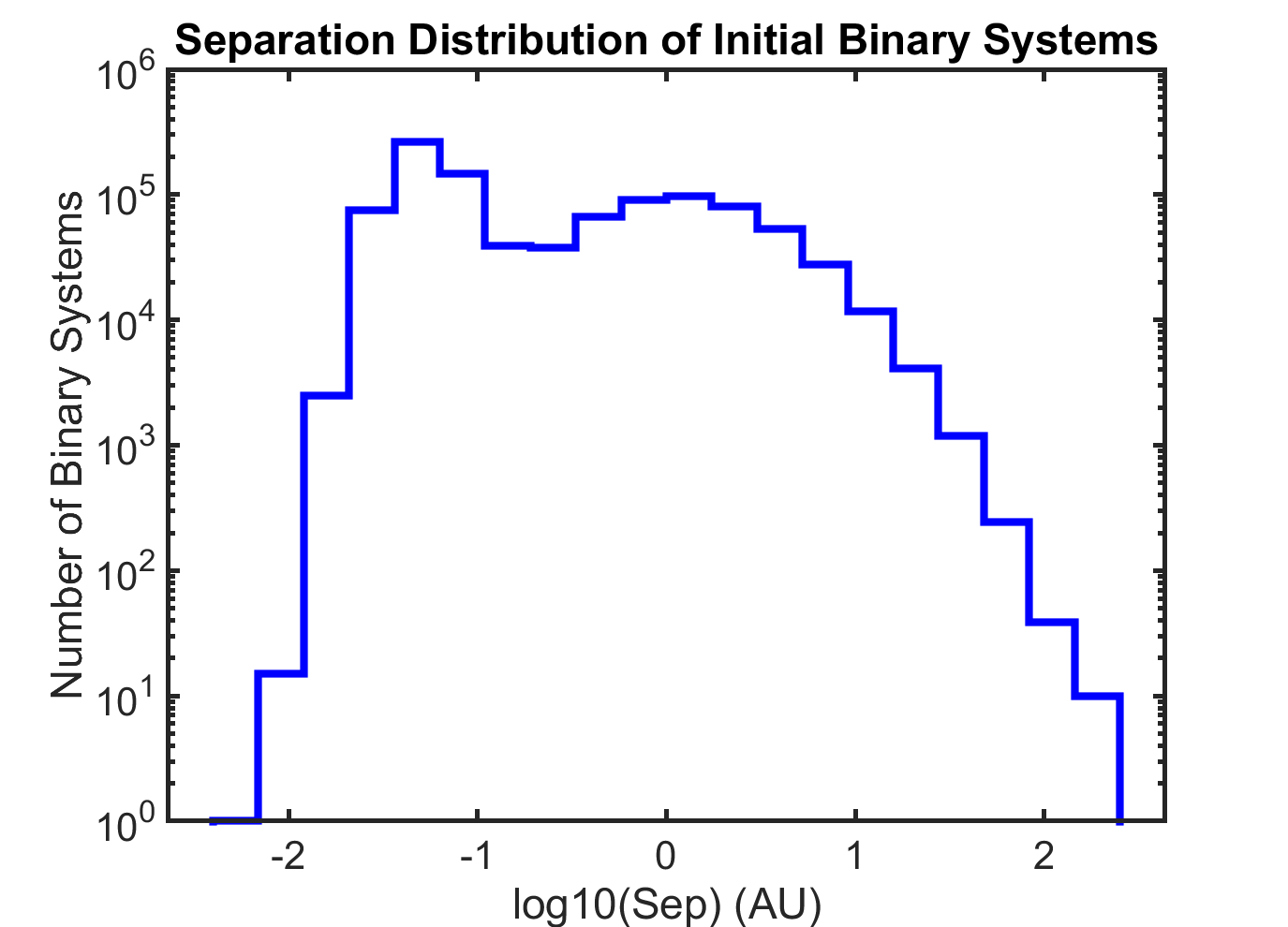}
      \label{fig: inisep}
    }
\caption{The initial distribution of $10^6$ systems of binaries that crossed their Roche limit after the EKL mechanism within the GC. Distribution properties were adopted  from \citet{Stephan+19}. The top panel shows the initial eccentricity (Red) which peaks around nearly circular and highly eccentric orbits. The bottom panels shows the initial separation (Blue) which displayed a wide range of distance from $10^{-2}$ AU to $10^{2}$ AU. 
}  
\label{fig:inihis}
\end{figure}

We use \citet{Stephan+16,Stephan+19} distribution, as stated above, at the onset of Roche limit crossing, to generate a population of $10^6$ binaries.
We show the  distribution of the systems' eccentricity and separation in Figure \ref{fig:inihis}. The two peak distribution of the SMA and eccentricity is expected due to the EKL and tidal evolution \citep[e.g.][]{Dan,Naoz+14,Rose+19}. 

We note that the additional distribution of wide binaries that do not cross their Roche limit may become unbound due to interaction with single stars in this dense environment \citep[e.g.,][]{Stephan+16,Rose+20}, and thus will not contribute to the compact object binary population. We also note that other formation channels such as BH-BH collisions \citep[e.g.,][]{O'Leary+09} or NS-BH capture \citep[e.g.,][]{Hoang+20} will have only small contribution to the overall population, but they explore different part of the parameter space (e.g., \S \ref{sec:2}). 

\begin{figure*}
    \centering
    \subfloat{
    \includegraphics[width=\columnwidth]{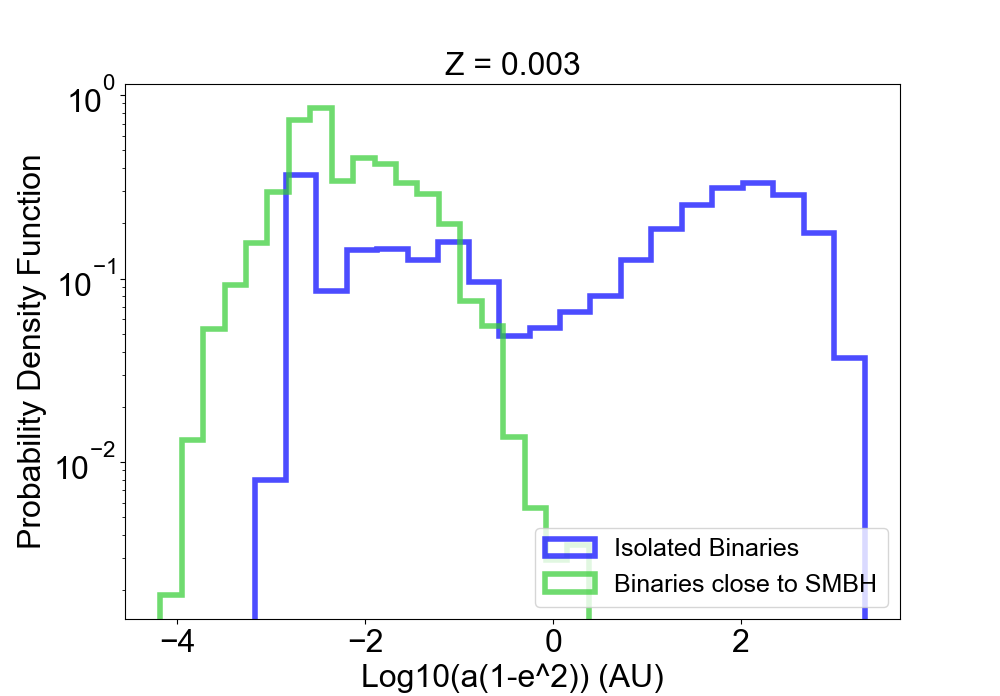}
    }
    \subfloat{
    \includegraphics[width=\columnwidth]{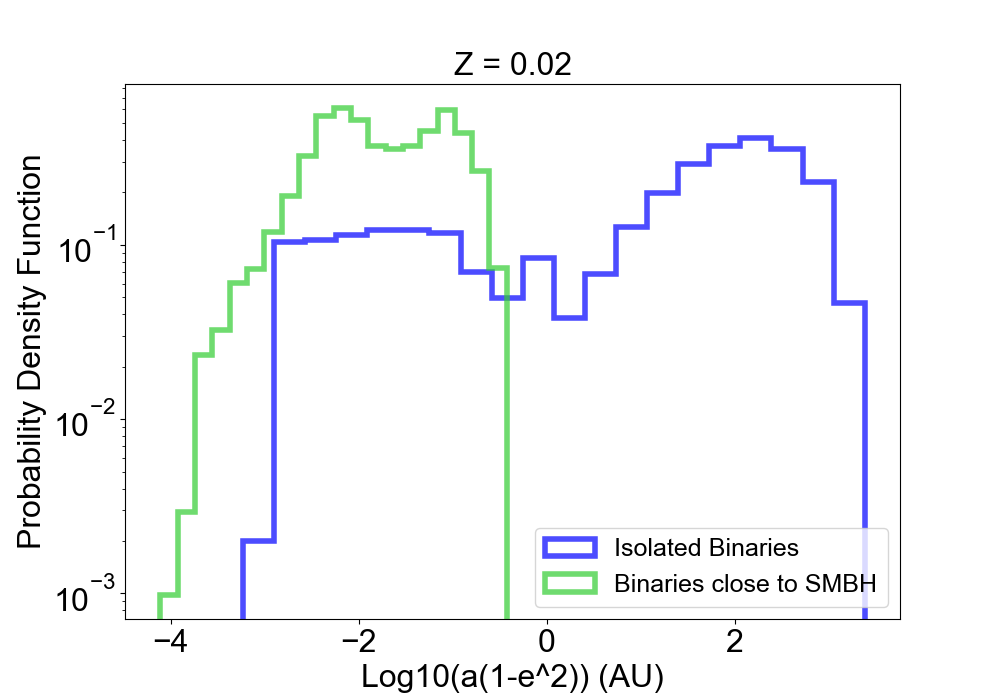}
    }\\
    \subfloat{
    \includegraphics[width=\columnwidth]{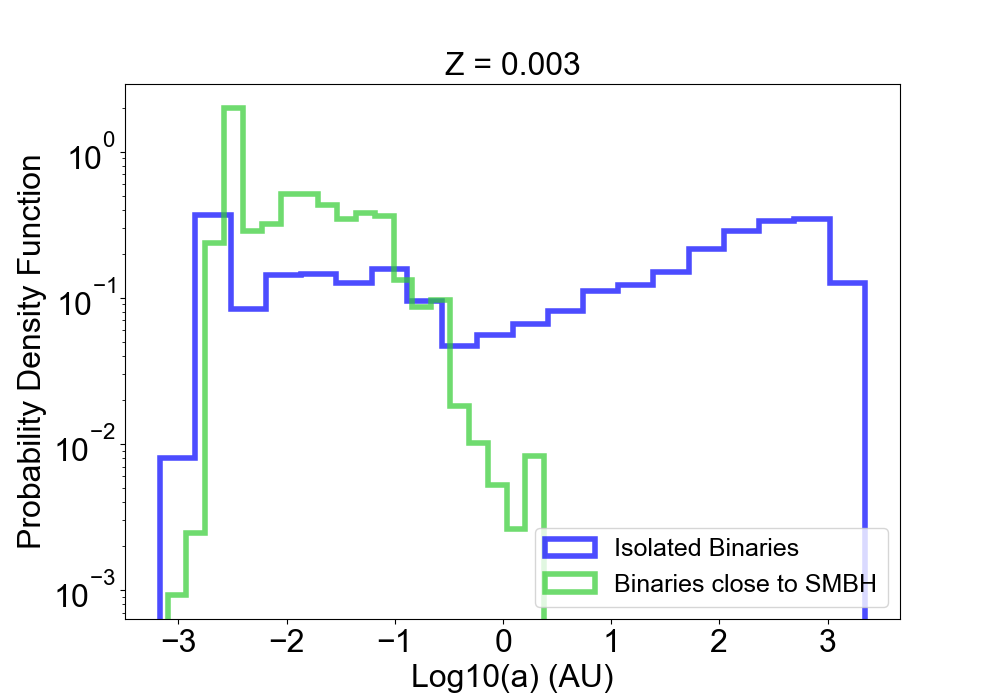}}
    \subfloat{
    \includegraphics[width=\columnwidth]{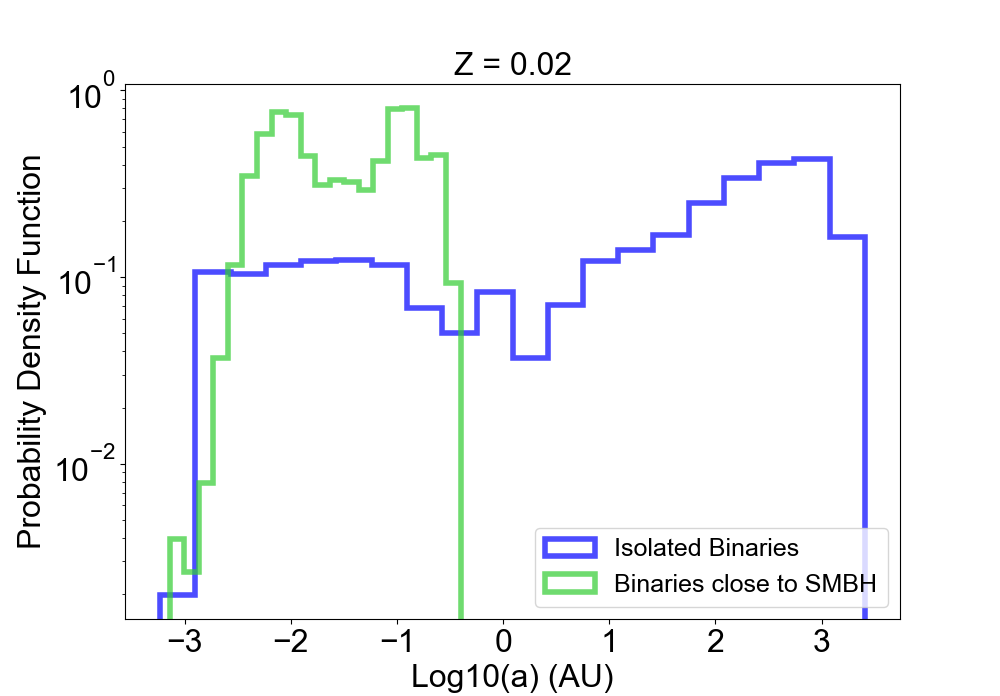}
    }
\caption{Distributions of Obrital Parameters for WD Binaries. We show the probability density functions of the specific angular momenta (top row) and semi-major axis values (bottom row) for both isolated binaries (blue) and binaries close to SMBHs (green). The systems are evolved with the stellar evolution code {\tt COSMIC}. Binaries close to SMBHs cannot be very wide due to stability consideration, thus limiting them to tight configurations and low specific angular momenta. }
\label{fig:unihis}
\end{figure*}


\subsection{Binary stellar evolution and compact object population}\label{sec:4.2}

As mentioned previously, {\tt COSMIC} is a stellar evolutionary code that not only follows individual stars beyond their main-sequence evolution after they exhaust their hydrogen fuel, but also interactions with a binary companion via mass transfer, common envelope.  \citet{Katie+19} presented the {\tt COSMIC} code, and evolved isolated stellar binaries all the way to their compact object stage. {\tt COSMIC} uses a modified version of {\tt BSE} which includes line-driven winds dependent on metallicities \citep[e.g,][]{Vink+01,Meynet+05,Grafener+08,Vink+11}, mass-loss transfer via Roche-limit overflow \citep[e.g.,][]{Hurley+02,Belzynski+08,Claeys+14}, and Supernova and natal kicks\citep[e.g.,][]{Fryer+01,Kiel+08,Fryer+12}, etc. 
The output of {\tt BSE} provides users the types of binary, binary state, separation, during each evolutionary stage. \citet{Katie+19} found around $10^8$ isolated compact binaries within Milky Way Galaxy as potential sources of electromagnetic and GW sources. Out of those, $10^4$ systems may be resolvable via LISA \citep{Katie+19}.

Since the publishing of \citet{Stephan+19}, \texttt{COSMIC} has been modified to include the effects of ultra-stripped SNe which experience reduced natal kick strengths due to the lack of a hydrogen envelope \citep{Tauris2015}. In comparison to standard core-collapse SNe, which have natal kicks drawn from a Maxwellian distribution with $\sigma=265\,km/s$ \citep{Hobbs2005}, \texttt{COSMIC} assumes natal kicks in ultra-stripped SNe are drawn from a Maxwellian distribution with $\sigma=20\,km/s$. This inclusion increases the local merger rate of BNSs such that it is consistent with the $90\%$ credible bounds of the LIGO/Virgo empirical rate \citep{Zevin2020}.

\begin{figure*}
    \centering
    \subfloat{
        \includegraphics[width=0.44\linewidth]{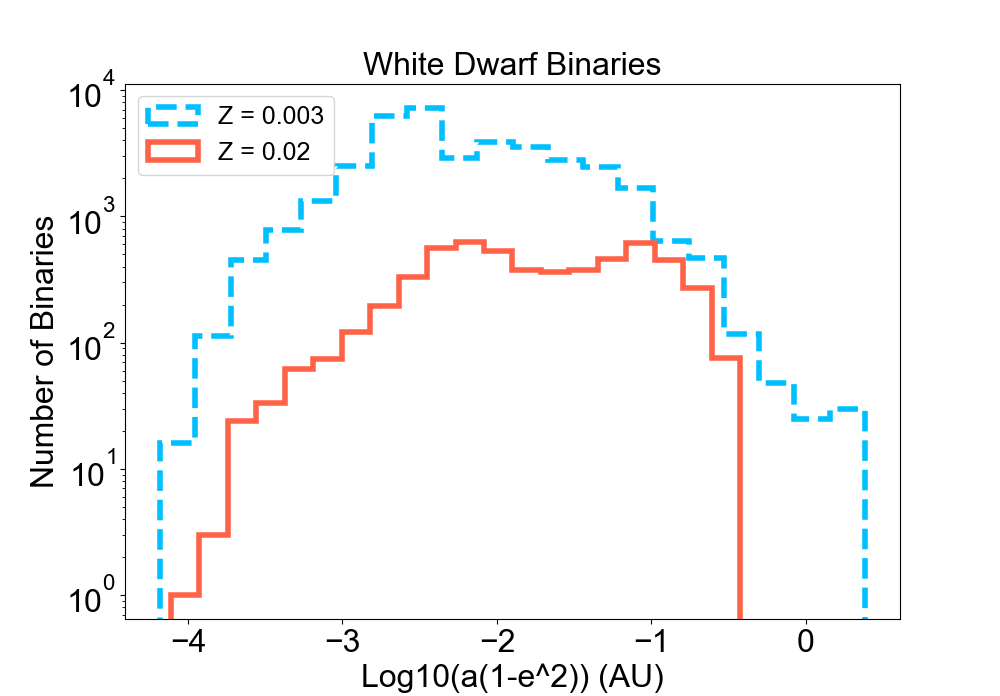}
    }
    \subfloat{
    \includegraphics[width=0.44\linewidth]{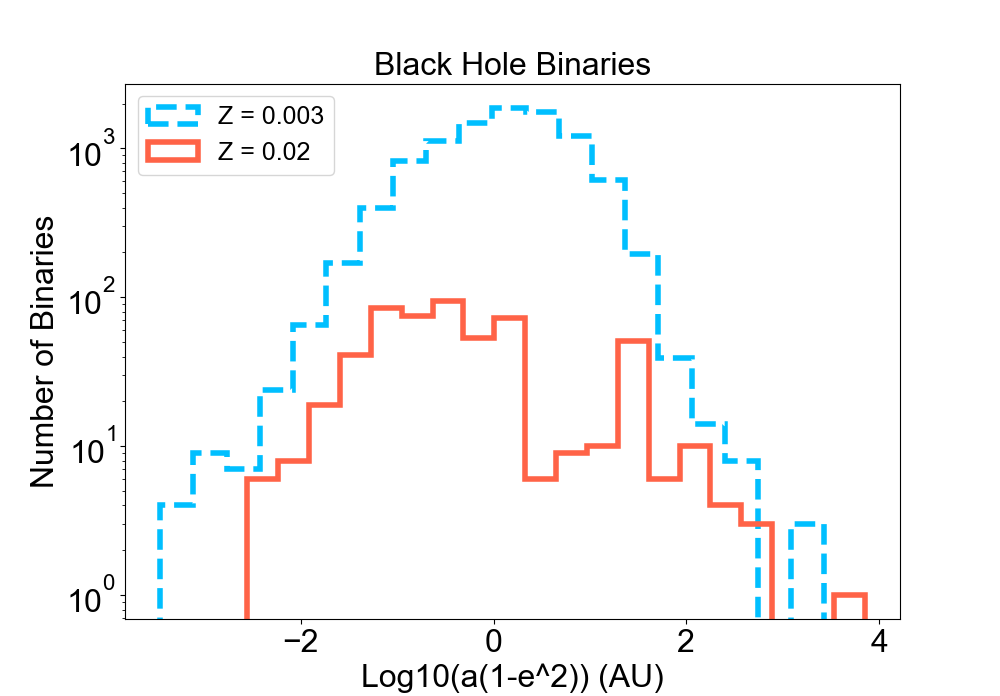}
    }\\
    \subfloat{
    \includegraphics[width=0.44\linewidth]{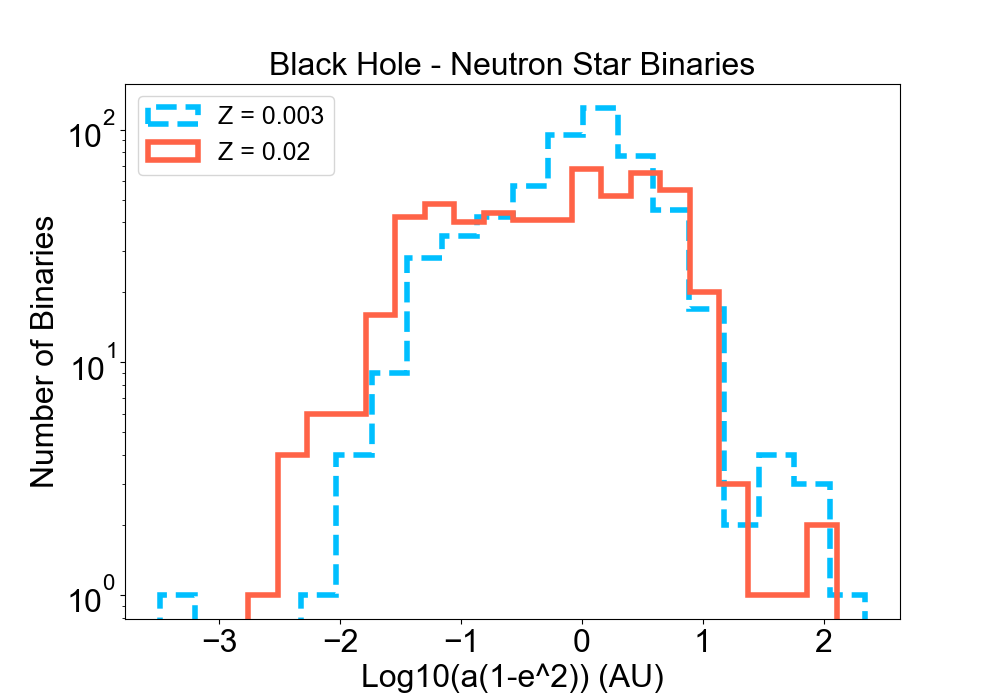}
    }
    \subfloat{
        \includegraphics[width=0.44\linewidth]{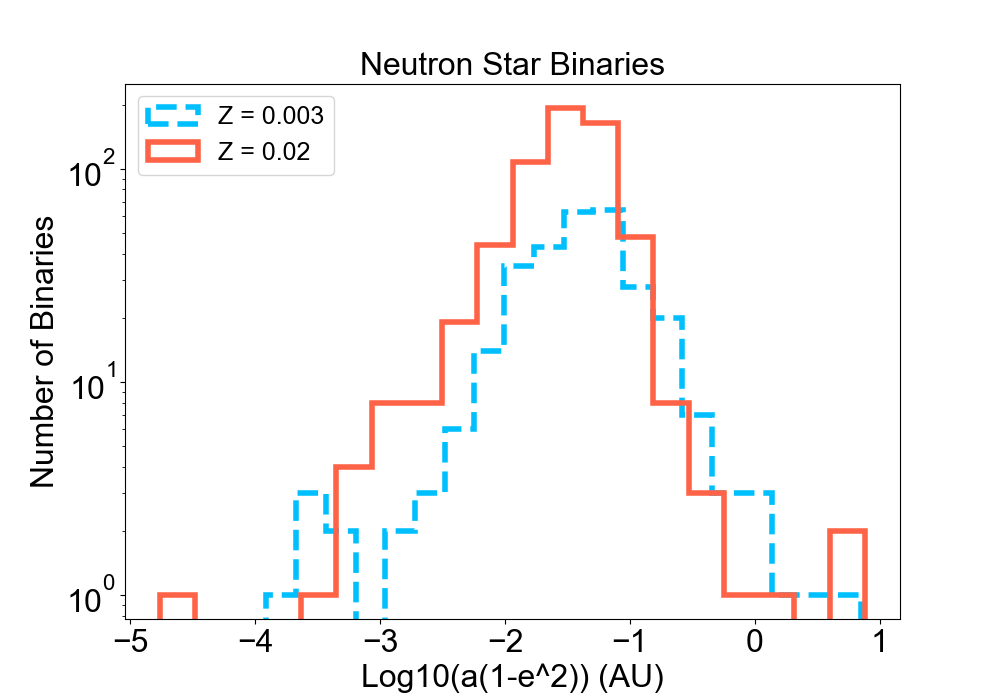}
    }\\
    \caption{The $10^6$ sets of initial binaries within the GC are evolved with  $Z = 0.02$ (red) and $Z = 0.003$ (blue). We select four types of compact object binaries. The x-axis takes the log scale of the specific angular momentum $a(1-e^2)$ while the y-axis is the number of binaries.  }
    \label{fig:meta}
\end{figure*}

As a first step, we investigate the effect our post EKL initial conditions have on the final compact object binary population. That is, comparing the evolution of isolated field binaries and binaries within the GC. Thus, we generate and evolve $10^6$  isolated stellar binary systems following \citet{Katie+19}.  We adopted an initial mass function from \citet{Kroupa+93}\footnote{We note that we use \citet{Kroupa+93} IMF here for consistency with \citet{Katie+19} results. The post-EKL calculations follow \citet{Stephan+16,Stephan+19}, which adopted Salpeter IMF. Since both IMFs drop in a similar manner for high mass stars, the effect is negligible \citep[e.g.,][]{Rose+19}.}, a thermal eccentricity model, and a log uniform orbital period. We assume all binary parameters are initialized independently and  assume a binary fraction of $50\%$.  \footnote{See \url{https://cosmic-popsynth.github.io/docs/stable/fixedpop/index.html} for more} All binaries are evolved for 13.7 Gyrs to capture their behavior over the age of the Universe. As a proof-of-concept we focus on WD binaries. The final distribution of those isolated WD binaries are shown in Figure \ref{fig:unihis} (blue).

We also considered the binaries that are on the onset of Roche limit crossing post EKL mechanism (e.g., Figure \ref{fig:inihis}). These binaries are evolved following the default settings defined in V3.3 of {\tt COSMIC} \citep{Breivik+20}. Note that high eccentricity is excited due to the EKL mechanism induced by the SMBH. Based on the expectation of the eccentricity distribution after EKL \citep[in the absence of tides e.g.,][]{Tey+13,Li+14,Naoz16,Rose+19},  we adopt a uniform distribution for binaries close to the SMBH as a proof-of-concept here.

In Figure \ref{fig:unihis} we focus on WD binaries, which represent the largest population in the system, and compare their population near an SMBH (green) to that of isolated binaries (blue, which their eccentricity does not excited). 

We show the semilatus rectum of populations of WD-WD, BH-BH, BH-NS, and NS-NS binaries for metallicities $Z = 0.02$ and $Z = 0.003$ in Figure \ref{fig:meta}. Filtering out all systems that have either unbound, merged, or filled their Roche limit, the percentage of each compact binary type evolved out of $10^6$ systems are described in Table \ref{ligo}. The discrepancies between our results with that of \citet{Stephan+19} are potentially due to different assumptions for wind and ultra-stripped SNe in {\tt COSMIC} as specified previously. In particular, the inclusion of ultra-stripped SNe explains the high abundance of NS-NS binaries in our simulations compared to \citet{Stephan+19}.

As apparent from Figure \ref{fig:meta}, metallicity significantly affects the abundance of compact objects. Metallicity as a function of stellar wind strongly affects the mass loss and final evolutionary outcome of massive stars \citep[e.g.,][]{Kudritzki+00,Nugis,Heger+03}. In particular, sub-solar metallicity yields a higher population for BH and WD binaries but a lower population for NS binaries.  For massive BH-BH binary progenitors, high metallicity enhances the solar wind, leading to smaller helium core, hydrogen envelope, and larger mass loss \citep[e.g.,][]{Langer+89,Hamann+95}. Hence, BH-BH binary progenitors with masses $> 40 {\rm M}_\odot$ and high metallicity might cause them to fallback  until only  neutron stars are made \citep{Heger+03}. For less massive stars ($\lesssim$ 10 ${\rm M}_\odot$) evolving into WDs, lower metallicity allows AGB stars to develop into more massive cores with higher luminosity, which increases the survival chances of a WD binary. In comparison, higher metallicity reduces the mass of helium core and increase the timescale for main sequence stars to form WD \citep[e.g.,][]{Heger+03,Yu+10,Romero+15}. For compact binaries involving NS (Bottom row of Figure \ref{fig:meta}), their population fluctuation is less sensitive to metallicity. This is due to the lower mass loss rate of their lower mass NS progenitors relative to BH progenitors. In addition, their natal kicks, which might unbind the binary, correlate little with their metallicity \citep[e.g.][]{Neijssel+19,Giacobbo+19}.

\subsection{LISA detections}\label{sec:4.3}

The post evolutionary results from {\tt COSMIC} presented here can be used in understanding the census of compact object binaries in the GC. Particularly, those compact object binaries have a chance to produce observable GW signals (see \S \ref{sec:2}). Those eccentric binaries emit GW signal that peaks approximately at 
\begin{equation}\label{fp}
    f_p (a,e) = \frac{(1 + e)^{1/2}}{(1-e)^{3/2}} \times f_{\rm orb}(a) \ ,
\end{equation}
where $f_{\rm orb}$ is defined in Equation (\ref{f_orb}) \cite[e.g.,][]{Hoang+19}. 
We assume that the outer semi-major axis of the triple hierarchy to be much wider than the inner orbit so that any GW back-reactions on the outer orbit are disregarded. In addition, we neglect oscillations in other orbital parameters due to EKL and GR since we expect them to only have peripheral effects on the $f_p$  and the modulation of the inclination $i$   and precession of pericenter are small near high eccentricity spike \citep[e.g.,][]{Naoz+13b}. $f_p$, combined with the chirp mass of the binaries, are used to explain the distribution of strain curvers within the LISA parameter space.

\begin{figure*}
    \centering
    \subfloat[]{
      \includegraphics[width=0.44\linewidth]{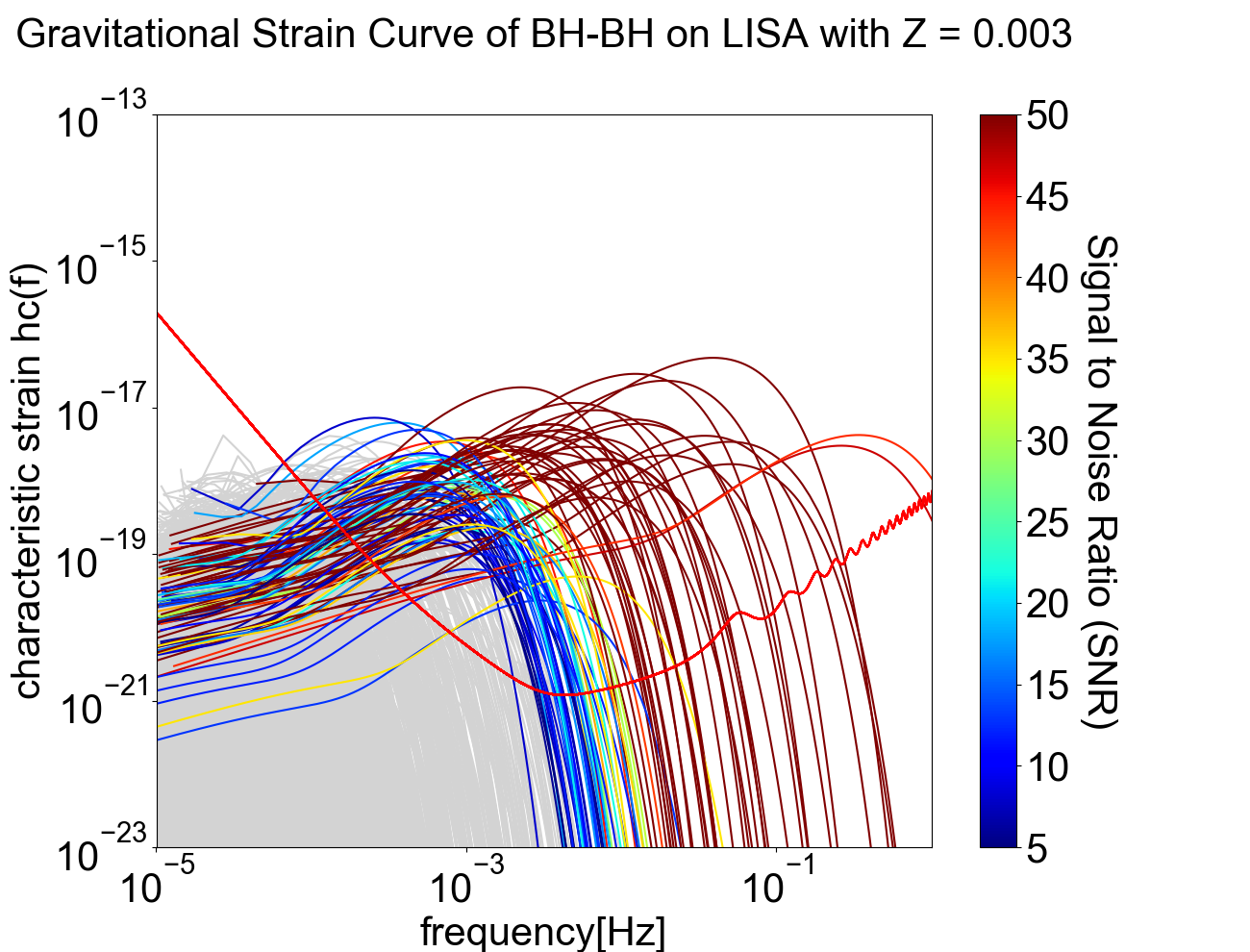}
        \label{fig:BH3}}
    \subfloat[]{
        \includegraphics[width=0.44\linewidth]{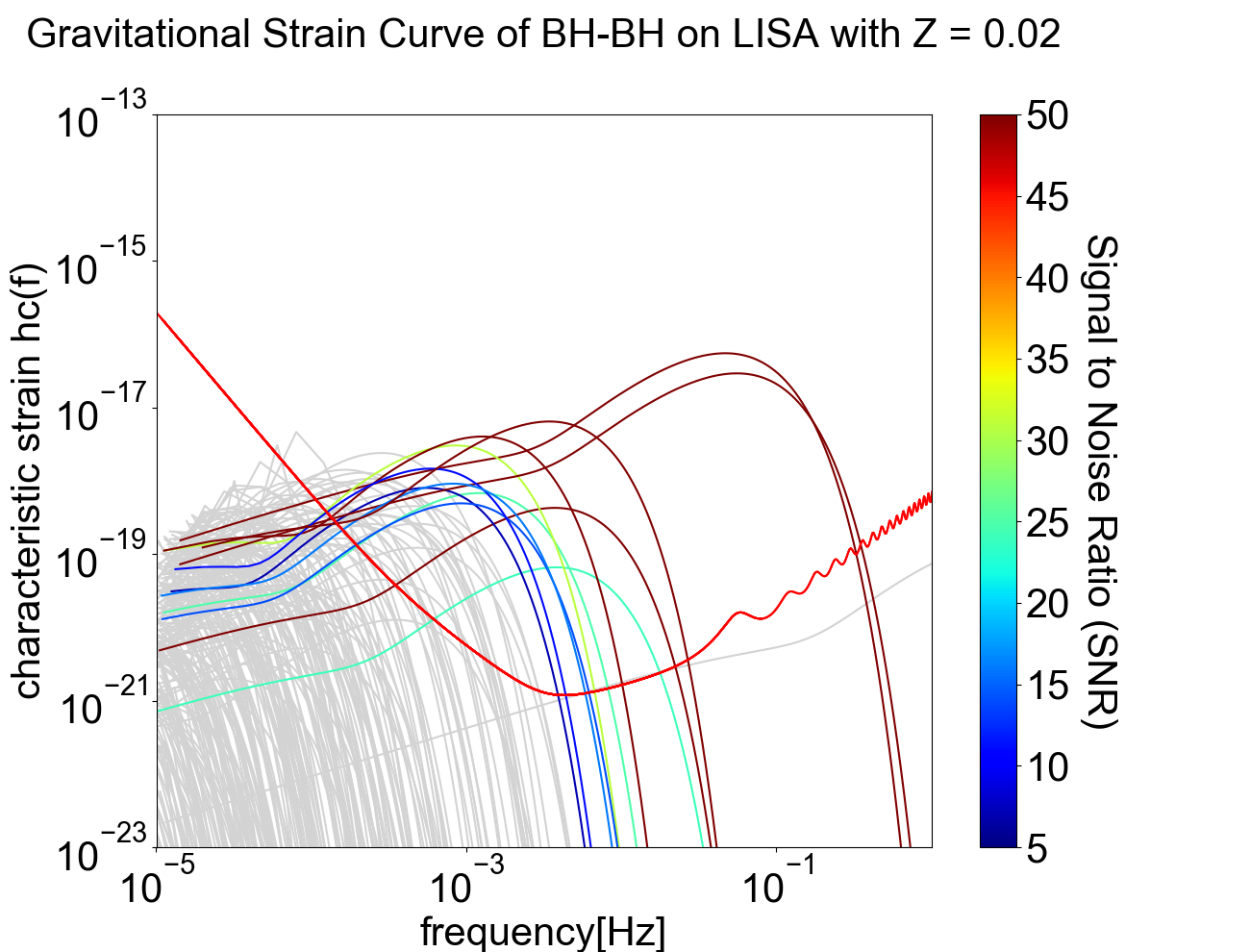}

        \label{fig:BH2}
    }
\caption{GW strain curve of BH binaries evolved via {\tt BSE} of {\tt COSMIC} with two metallicities  $Z = 0.02$ (right) and $Z = 0.003$ (left). A uniformly distributed $e$ is taken from $0$ to $1$ for all binaries. The strain curves above the LISA sensitivity curve (red) are color-coded according to the each Signal to Noise Ratio (SNR), all greater than 5. The grey curves in the background are systems with SNR $< 5 $. }
\label{fig:BH}
\end{figure*}

\begin{figure*}
    \centering
    \subfloat[]{
      \includegraphics[width=0.48\linewidth]{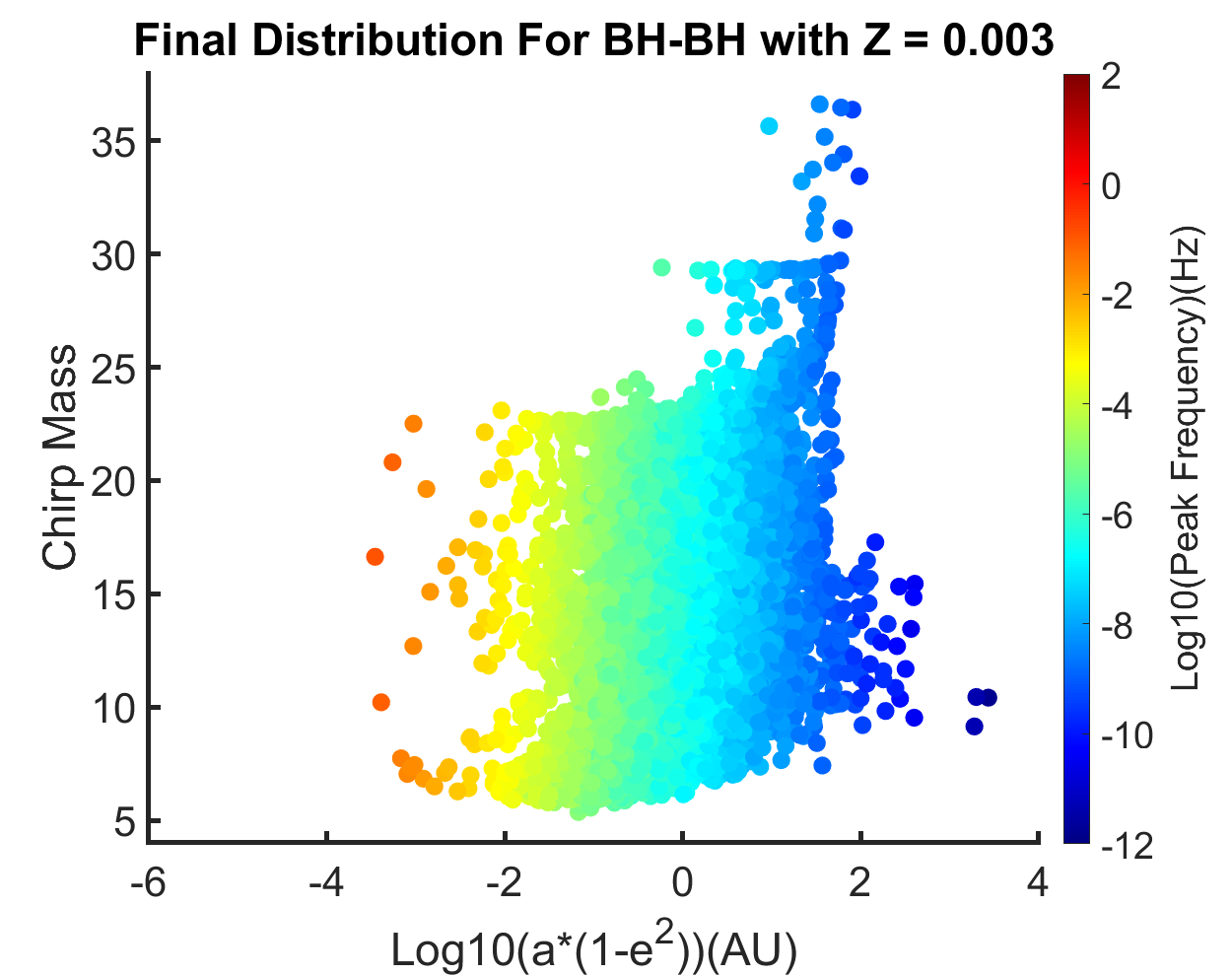}
        \label{fig:scatter3}
    }
    \subfloat[]{
      \includegraphics[width=0.48\linewidth]{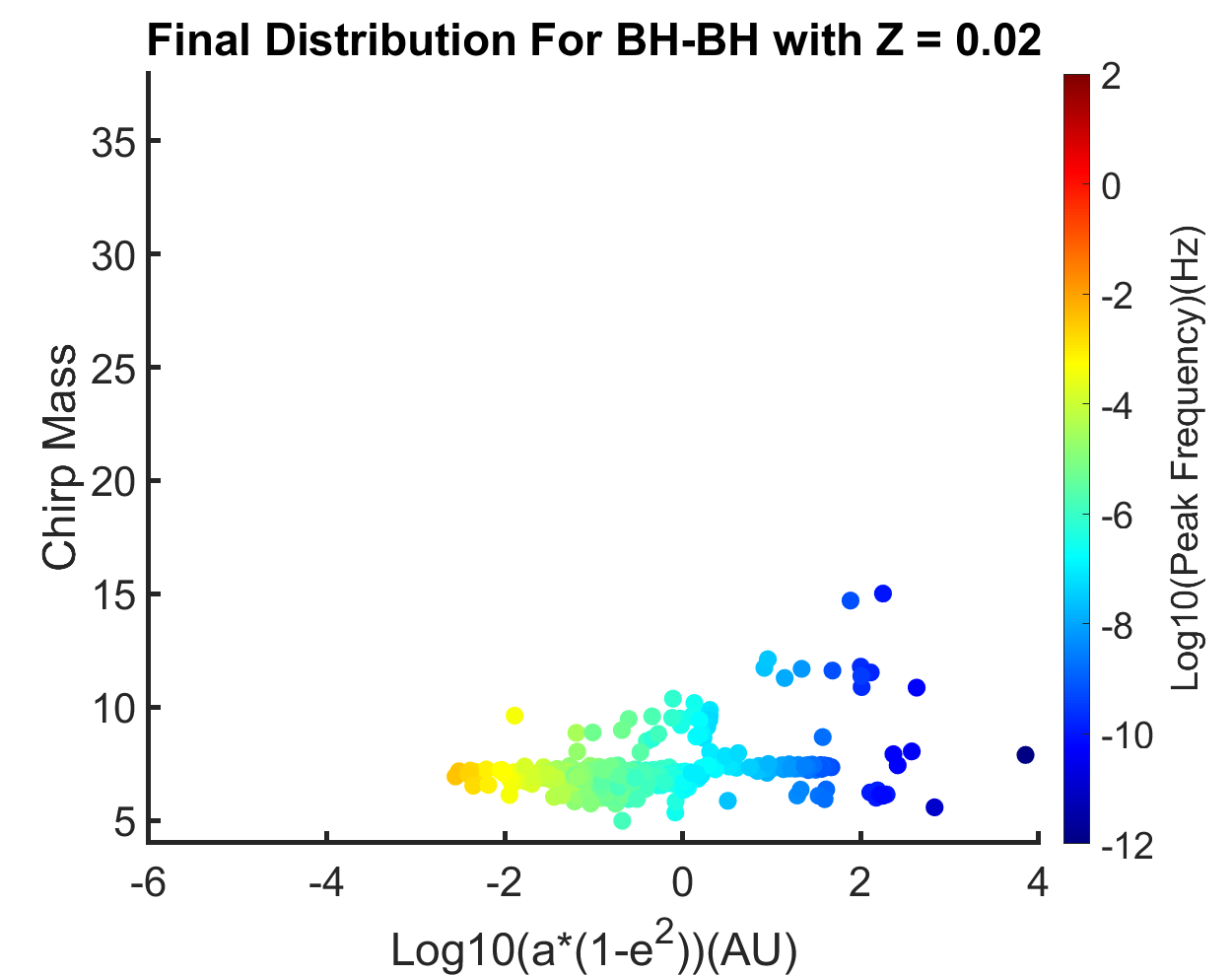}
        \label{fig:scatter2}
    }
\caption{The population distribution of BH binaries evolved via {\tt COSMIC} with $Z = 0.003$ (left) and $Z = 0.02$ (right). The x-axis gives the specific angular momentum, log10$(a(1-e^2))$, while the y-axis is the chirp mass of the binary system given in Equation (\ref{cm}). The scattered points are color coded according to their peak frequency given in Equation (\ref{fp}). The lack of massive chirp mass BH binaries within the higher peak frequency region of the higher metallicity sample (right) provides insight to the lack of visible binary systems in Figure \ref{fig:BH2}. }
\label{fig:scatter}
\end{figure*}

\begin{figure*}
    \centering
    \subfloat[]{
      \includegraphics[width=0.47\linewidth]{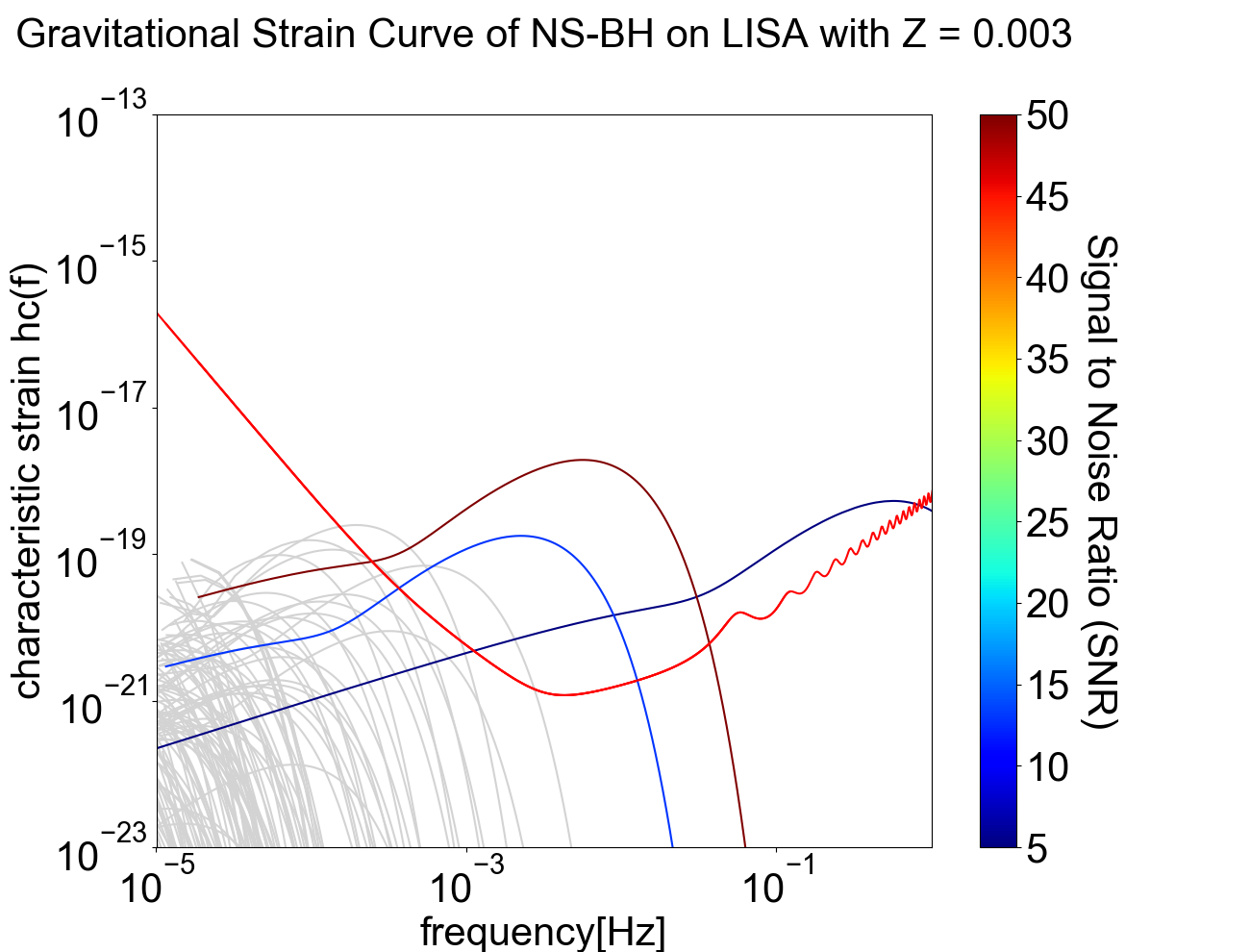}
    \label{fig:NSBH3}
    }
    \subfloat[]{
      \includegraphics[width=0.47\linewidth]{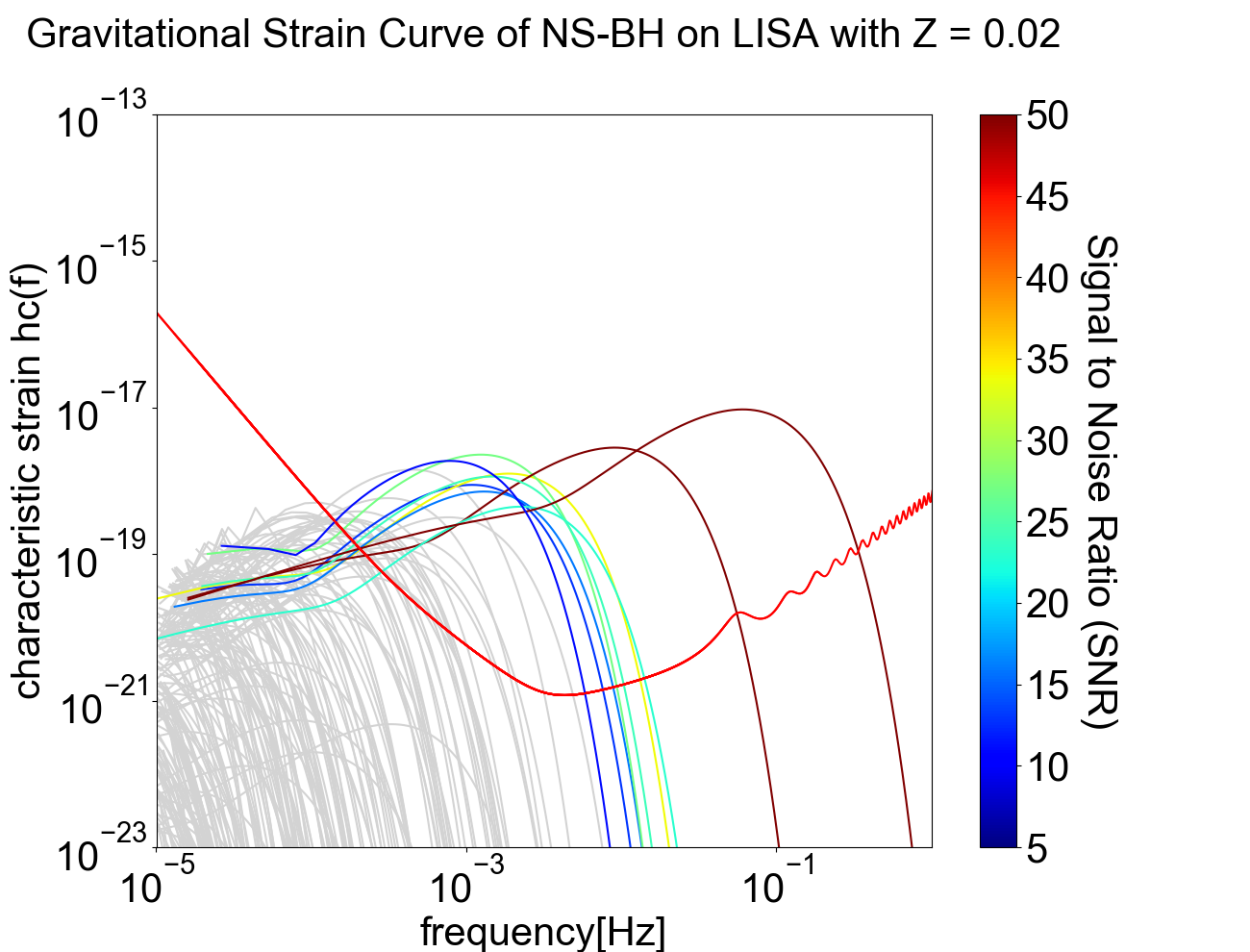}
     \label{fig:NSBH2}
    }\\
    \subfloat[]{
      \includegraphics[width=0.47\linewidth]{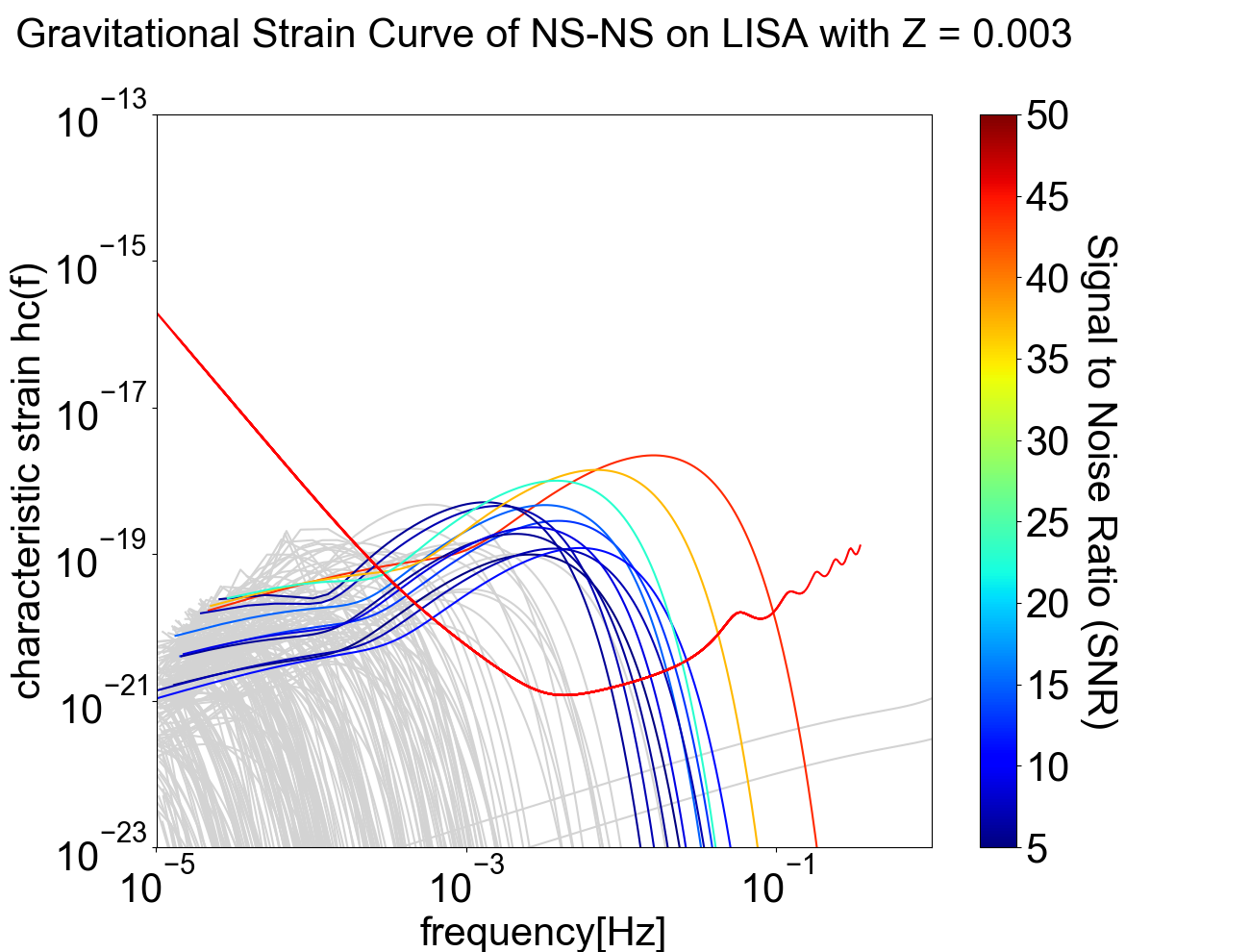}
      \label{fig:NS3}
    }
    \subfloat[]{
      \includegraphics[width=0.47\linewidth]{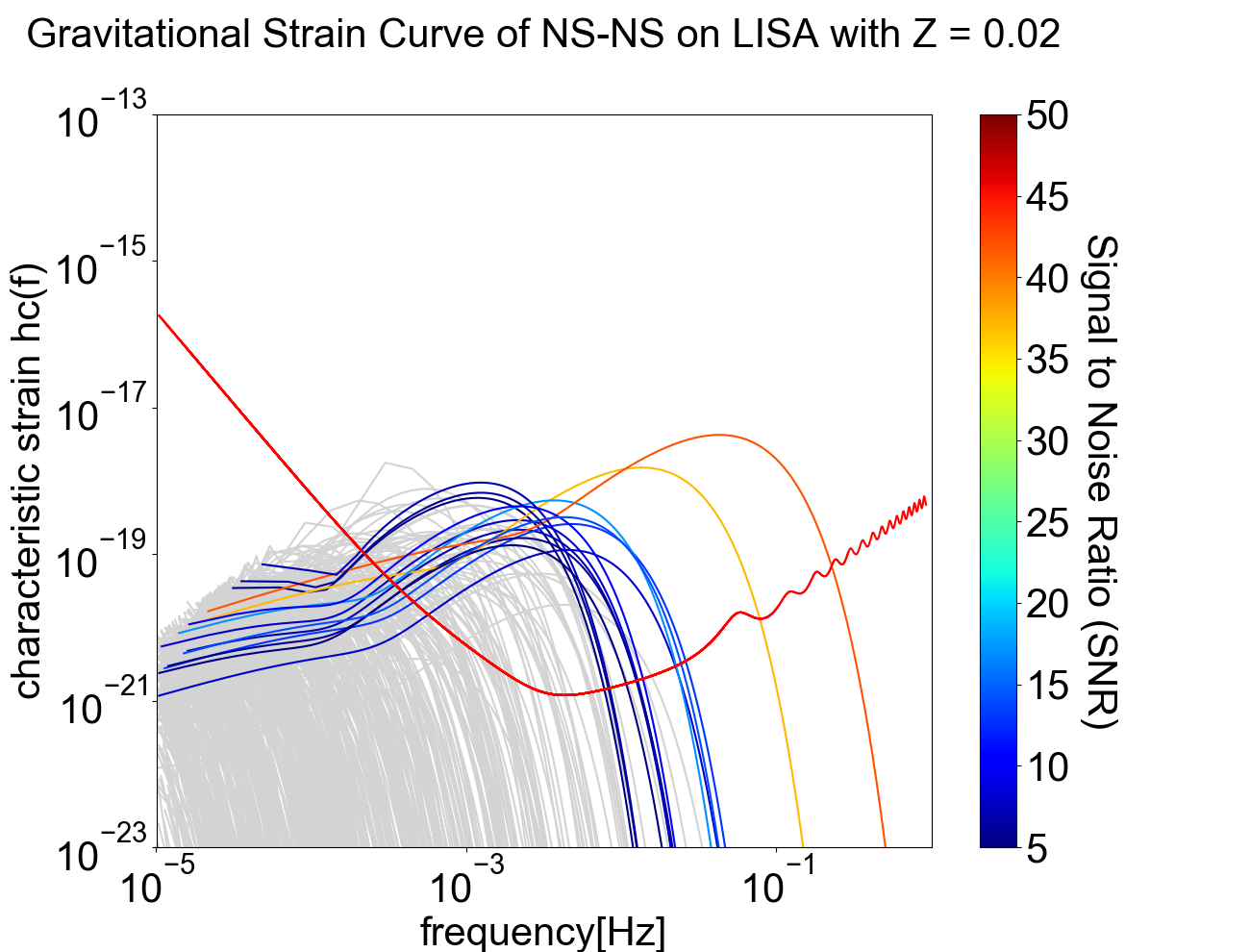}
        \label{fig:NS2}
    }\\
    \subfloat[]{
      \includegraphics[width=0.47\linewidth]{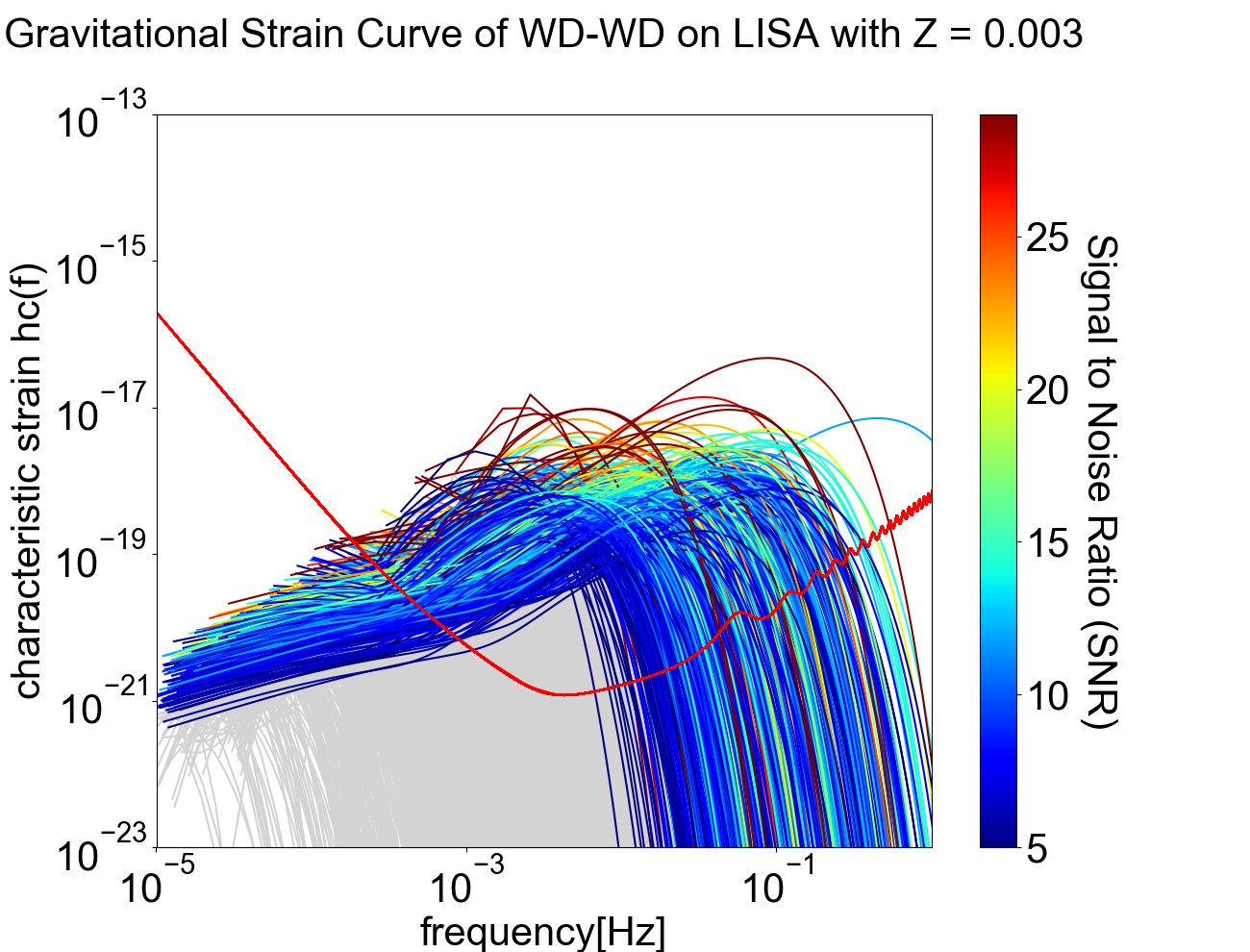}
      \label{fig:WD3}
    }
    \subfloat[]{
      \includegraphics[width=0.47\linewidth]{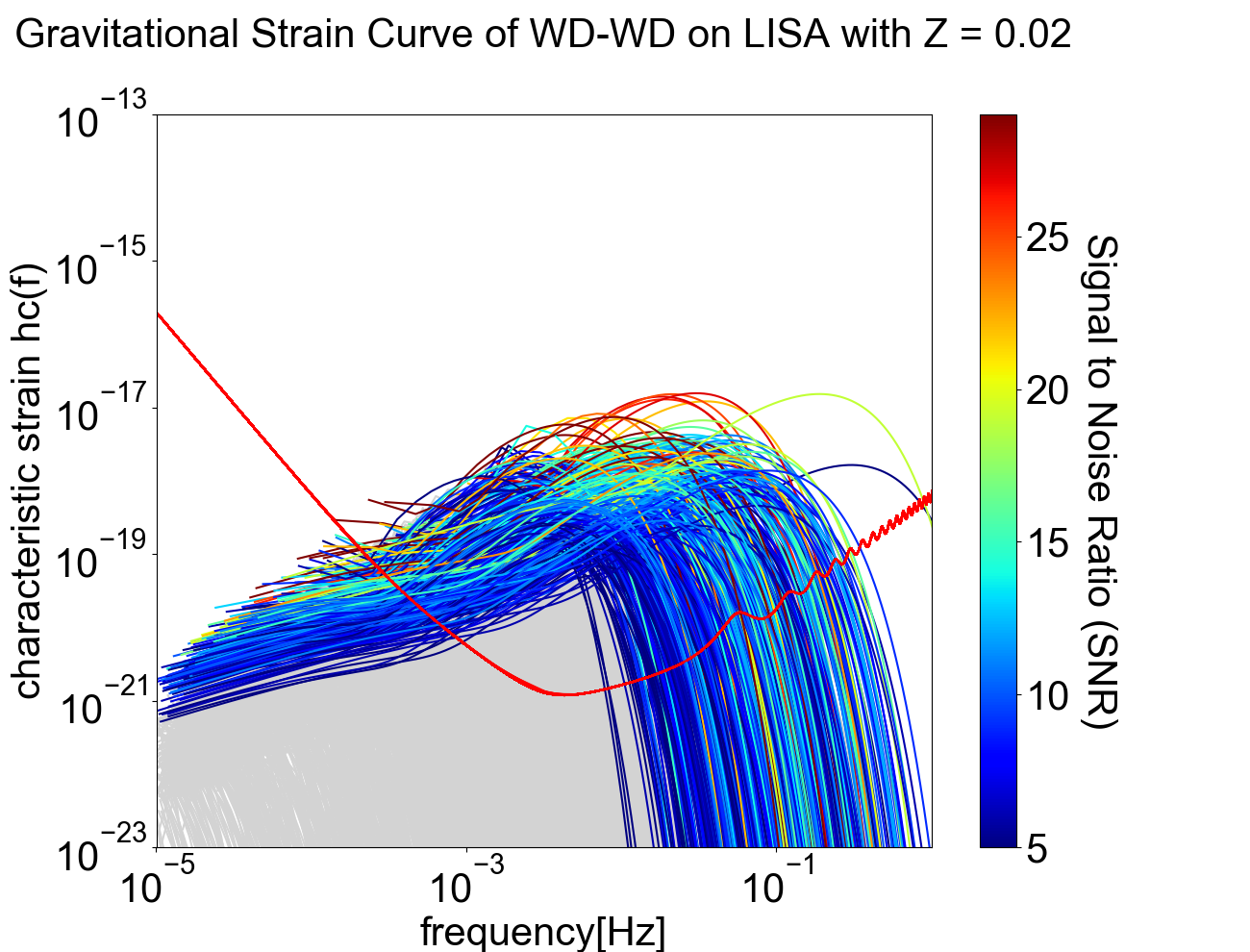}
      \label{fig:WD2}
    }\\
\caption{Strain Curves of NS-BH binaries (top row), NS binaries (middle row) and WD binaries (bottom row) evolved via {\tt COSMIC} with  $Z = 0.02$ (right column) and $Z = 0.003$ (left column). All other initial conditions follow Figure \ref{fig:BH}.
}
\label{fig:NS}
\end{figure*}

As mentioned before, since the inner orbit of the triple hierarchy system is expected to undergo the EKL mechanism, we re-assign a uniform distribution of eccentricity between $0$ and $1$ to the output binaries of {\tt COSMIC}. The luminosity distance is set to be $D_l = 8 {\rm kpc} = 1.65 \times 10^9$ AU. As previously discussed, we adopt an observation time which is the minimum of LISA's $4\,\rm{yr}$ mission duration or the binary's merger timescale. Here we explore the two evolved systems of binaries via {\tt COSMIC} in \S \ref{sec:4.2} with $Z = 0.02$ and $Z = 0.003$. The detectability of each system is visualized by overplotting each system's GW strain curve, squareroot of Equation (\ref{strain}), onto the LISA sensitivity curve (Red, Figure \ref{fig:BH} and \ref{fig:NS})\footnote{The detailed equations for constructing the LISA sensitivity curve can found in \citet{Robson+19}'s paper.}. We note that out of $10^6$ binaries, no  binary with SNR$>5$, of any types in either metallicity, merges within $10$~years, which renders the SNR calculation from Equation (\ref{eq:tscale}) an appropriate approximation.

We note that previous studies on compact object binaries population such as \citet[]{Yu+10,Belczynski+10,Liu+14,Lamberts+18,Katie+19,Lau+20,Sesana+20} have focused on simulating isolated binary sources of LISA in the Milky Way Galaxy including disk, the bulge, and halo \citep[e.g.,][]{Nelemans+01b,Nissanke+12,Korol+17}, while \cite{Kremer2018} study the population from Milky Way globular clusters. Here our simulation of eccentric compact object binaries sources within the GC provides similar yet distinctive results for the population potentially detectable by LISA. Since the eccentricities of each source changes under the EKL mechanism, we expect the actual number of observable sources to fluctuate up and down slightly from our current result. 

In Figure \ref{fig:BH}, we consider BH-BH binaries with the orbital parameters shown in Figure \ref{fig:meta}, and color code the SNR, with grey being below $5$. The main difference between the solar and sub-solar metallicity is the number of systems that are potentially detectable, where the sub-solar calculation yields larger abundance of detectable systems. This result is consistent with isolated binaries in the field \citep[e.g.,][]{Katie+19}, and can be further understood with the aid of Figure \ref{fig:scatter}. Figure \ref{fig:scatter} depicts the chirp mass, defined as:
\begin{equation}\label{cm}
    \mathcal{M} = \frac{(m_1 \times m_2)^{3/5}}{(m_1 + m_2)^{1/5}} \ ,
\end{equation}
 plotted against the specific angular momentum of the systems. The  mass of a system, which determines the overall amplitude of its GW signal ($h_0$, Eq.~(\ref{eqn:ho})), is visualized via the chirp mass. As shown in Figure \ref{fig:scatter} and Table \ref{ligo}, not only the overall population of the solar metallicity systems are $5\%$ out of that of the sub-solar ones, but they have smaller chirp masses. In other words, the sub-solar metallicity systems have higher BH masses that result in higher peak frequency, thus rendering them in the LISA detectable region.

For NS-BH systems (top row in Figure \ref{fig:NS}), we found that the total amount of systems did not change much between the two metallicity values we adopted (e.g., Figure  \ref{fig:meta}). But due to the larger chirp mass and lower orbital frequency their overall GW amplitude is smaller. Note that if NS-BH is formed via GW capture \citep[e.g.][]{Hoang+20}, their separation will be much smaller than the post EKL and binary stellar evolution prediction that is presented here. Therefore these systems may still be detectable (see Figure \ref{fig:snr}). Additionally, stronger natal kicks in higher metallicity might lead to lower separation \citep[e.g.][]{Lu+19}, which results in more visible systems then presented in Figure \ref{fig:NSBH2}.

For NS-NS systems, the simulated observable systems have SNR closer to $10$ in either metallicity (Figure \ref{fig:NS} middle row). While more systems are observable with higher SNR compared to NS-BH systems, they are less abundant than the BH-BH systems. This behavior is due to the concentration of systems within the low frequency region ${\rm log} f_p < -4$, of which only higher eccentricity pumps due to EKL might render some systems observable. Furthermore, note that as can be seen in Figure \ref{fig:meta}, the solar and sub-solar metallicity differ only by about a factor of $2$, resulting in a similar signal.

Maybe the most interesting feature we find is for WD-WD systems, depicted in Figure \ref{fig:NS} bottom row. First, it is worth noting the WD binaries are predicted to be at the highest abundance. That is not surprising due to the initial mass function (IMF) of saltpeter adopted by \citet{Stephan+16,Stephan+19}. In this IMF, the low mass stars are at high abundance, which is also consistent with \citet{Stephan+19} results. Additionally, the semi-major axes of the WD binaries is relatively short (as can be seen in Figure \ref{fig:inihis}), which raise their orbital frequency to that of LISA (log $f_p > -3$). Lastly, as mentioned above, we adopt a uniform eccentricity distribution which yields a smaller specific angular momentum. Therefore the combination of all of these factors results in a large abundance of single detections of WD binaries at the galactic center. 

We also offer an approximation for the number of visible sources per galaxy as,
\begin{equation}
\begin{split}
    {\rm N} =   N_{\rm steady\_state} \times f_{\rm Roche} \\ \times  f_{\rm SNR > 5}  \times f_{\rm binary\_fraction}
\end{split}
\end{equation} \, 
In which, 
\begin{equation}
     N_{\rm steady\_state} = {\rm SFR} \times t_{\rm steady\_state}
\end{equation}

where we assume a star formation rate\footnote{Here we approximate the star formation rate as the binary formation rate because we are interested in massive stars and the majority of them if not all of them are in binaries \citep[e.g.,][]{Raghavan+10a,Sana+12,Moe+17}}  of SFR$=10^{-3}/{\rm yr}$ and take $f_{\rm Roche} \sim 0.25$ as the fraction of binaries that merge via Roche limit crossing \citep{Stephan+19}. Additionally, $N_{\rm steady\_state}$ is the number of stars in stead-state at the galactic nuclei. Motivated by our own galaxy, we adopt $N_{\rm steady\_state} \sim 10^5 - 10^6$. Lastly, $f_{\rm binary\_fraction}$ is the fraction of each type of binaries specified in column 3 of Table \ref{ligo}, and $f_{\rm SNR > 5}$ is the fraction of visible binaries in LISA out of each type of compact object binary. Using this estimation we find that per galaxy, EKL mechanism may yield about 14 - 150, 0.02 - 2, 0.2 - 4, and 0.25 - 20 of WD-WD, NS-BH, NS-NS, and BH-BH binaries that could be visible within the LISA band (The full breakdown can be found in Table \ref{ligo}.)


\subsection{LIGO detections}\label{eligo}
\begin{figure}
  \centering
  \includegraphics[width=\linewidth]{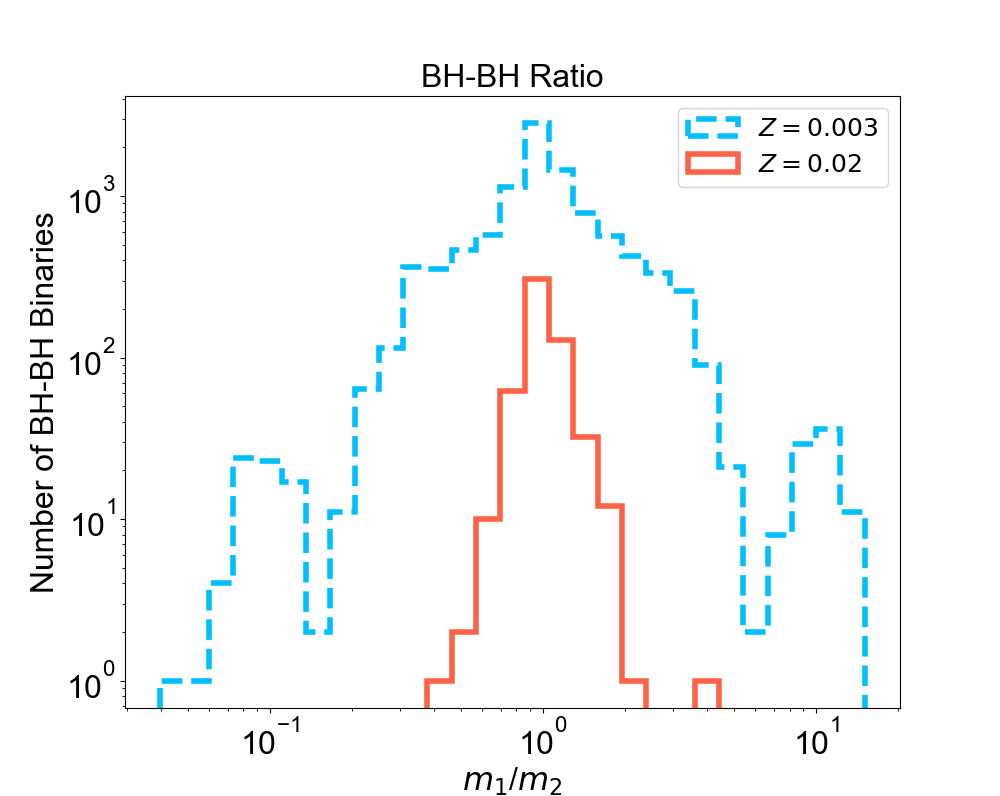}
\caption[]{The mass ratio of BH-BH binaries of $Z = 0.02$ (red) and $Z = 0.003$ (blue) post {\tt COSMIC} evolution. Both the y-axis and the bins are in log scale. The standard deviation of mass ratio in log scale are $\sigma_{0.02} = 0.19$ (red) and $\sigma_{0.003} = 0.61$ (blue) respectively. $\dagger$ We note that here we present the x-axis as the mass ratio of the initial conditions of each system instead of conforming to the usual choice of $q$}
\label{fig:BHratio}

\end{figure}

\begin{table}
\caption{The third column presents the fraction of compact object from $10^6$ systems after evolving with {\tt BSE} code in {\tt COSMIC}.  The fourth column is the EKL {\it induced LIGO detection rate} of those sources with $f_{\rm SMBH} = 1$ and SFR = $10^{-3}/$yr. The fifth column gives the estimated number of detectable events within LISA band during its lifetime per metallicity. $^\dagger$We note that the actual merge rate may be large because hard binaries will merge simply due to GW emission, with no need for EKL assitance. The last column presents the number of detectable sources of compact binary per galaxy via LISA. $^*$We note that WD-WD merges are not visible in LIGO/Virgo Detector.  }
\centering
\label{ligo}
\begin{tabular}{lllll}
\noalign{\smallskip} \hline \hline \noalign{\smallskip}
$Z$ & Population & Fraction & EKL Merger & LISA Band  \\
& & & Rate$^\dagger$ & Visibility\\
 &    & ($\%$)  & (${\rm Gpc}^{-3}/{\rm yr}$) & (\#)
\\
\hline
$0.003$ & BH-BH & 1.00 & 5 & 2 - 19 \\ 
& NS-NS & 0.03 & 0.15 & 0.2 - 2 \\
& NS-BH & 0.05 & 0.25 & 0.025 - 0.25  \\
& WD-WD* & 4 & 19 & 15 - 148\\
\noalign{\smallskip} \hline \noalign{\smallskip}
$0.02$ & BH-BH & 0.06 & 0.3 &  0.3 - 3 \\ 
& NS-NS & 0.06 & 0.3 & 0.4 - 4\\
& NS-BH & 0.06 & 0.3 & 0.2 - 2 \\
& WD-WD* & 3 & 15 & 14 - 137 \\
\noalign{\smallskip} \hline \noalign{\smallskip}
\end{tabular}
\end{table}

Additionally, massive mergers such as BH-BH and NS-BH are also potential signal sources of LIGO \citep[e.g.][]{Fragione+19a,Fragione+19c,Fragione+19b}. 
To allow for quantitative comparison with the results of \citet{Stephan+19}, we adopted their parameters. In what follows, we assume a similar galactic condition as that of Milky Way to estimate potential signals from compact object binaries of other galaxies via LIGO up to $\sim 1.5$ Gpc \citep{abbott+18}. We assume a galaxy density of $0.02$~${\rm Mpc}^{-3}$ \citep{Conselice+05} and a star formation rate of SFR$=10^{-3}/{\rm yr}$. The latter is estimated based on the Milky Way properties \citep[e.g.][]{Genzel+03,schodel+03}. The fraction of galaxies with a SMBH in the center ($f_{\rm SMBH}$) is approximated between $1-0.5$ \citep[e.g.,][]{Ferrarese+05,Kormendy+13}, where we adopt a unity value. Additionally, \citet{Stephan+19} predicted that the fraction of binaries that will merge via crossing the Roche limit is $f_{\rm Roche} \sim 0.25$. 
Combining, the LIGO detection rate, $\Gamma$, of each type of binary within $1$~${\rm Gpc}^3$ via EKL merger channel is, \begin{equation}\label{rate}
\begin{split}
        \Gamma = \left(\frac{0.02 \times {\rm Galaxies}}{\rm Mpc^3}\right) \left( \frac{10^9 \times {\rm Mpc}^3}{{\rm Gpc}^3} \right) \times f_{\rm SMBH} \\\times {\rm SFR} \times f_{\rm Roche/merger} \times f_{\rm EKL} \times f_{\rm binary\_type} \, 
\end{split}
\end{equation}
where, $f_{\rm EKL}$ is the fraction of the compact object binaries that will merger due to EKL, conservatively estimated as $f_{\rm EKL}\sim 0.1$ \citep[e.g.][]{Hoang+18}. The binary type, $f_{\rm binary\_type}$ are specified in the third column of Table \ref{ligo}. We provide the summary of the possible detectable events in the fourth column of Table \ref{ligo}. For the BH binaries, those results are within a factor of $\sim 3$ to $4$ from \citet{Stephan+19}'s result using $Z = 0.02$ but fairly consistent with their estimates for NS-BH. As mentioned above this is a result of the updated version of {\tt COSMIC} and the assumptions of the   natal  kicks  in  ultra-stripped SNe.  We note that in \citet{Stephan+19}, there were no NS-NS present, but with this updated {\tt COSMIC} package, see above, more NS-NS are present.  In addition, the binary stellar evolution prescription here creates a large uncertainty when estimating the possible rates.

In Figure \ref{fig:BHratio} we depict the distribution of BH binaries mass ratio post {\tt COSMIC} evolution. As shown in this figure, both distributions have wings of mass ratio deviating from $1:1$, with the sub-solar population possessing a larger variance. Interestingly, previous LIGO/Virgo have found a mass distribution which indicated similar mass components \citep[e.g.][]{Abbott+16a,Abbott+16b,Abbott+16c,Abbott+16d,Abbott+17a,Abbott+17b,Abbott+17d,Abbott+17c,Abbott+19b,Abbott+20c,Abbott+20a}. However, two events GW190412 and GW190814 \citep[e.g.,][]{Abbott+19c,Abbott+20b} possess mass ratio that differs significantly from unity. As mentioned above, the mass ratio distribution in our model, following \citet{Stephan+19} setting, was taken from a Gaussian distribution with a mean of $0.23$ and a standard deviation of $0.42$. Thus, while the deviation from the mass ratio of unity is an imprint of the initial conditions, it is interesting to note that the sub-solar metalicity population has larger range of mass ratios.

\section{Discussion} \label{sec:dis}

 Almost every galaxy, including Milky Way, hosts a SMBH in its center surrounded by crowded nuclear stellar clusters. The proximity of the Milky Way's SMBH provides a unique opportunity to explore dynamics and phenomena that ought to exist in other galaxies. It has been suggested that stellar binaries are of high abundance around the nuclear cluster of GC \citep[e.g.,][]{Ott+99,Martins+06,Pfuhl+14,Stephan+16,Stephan+19}. Within  the vicinity of a SMBH a binary has to be on a tighter configuration compared to its orbit around the SMBH, and hence undergoes the EKL mechanism \citep[e.g.,][]{Antonini+10,Antonini+12,Prodan+15,Stephan+16,Stephan+19,Hoang+18}. Recently,  \citet{Stephan+19} showed that combining EKL evolution with single and binary stellar evolution yields a high abundance of compact object binaries and has the potential to become GW sources for the LIGO/Virgo and  LISA. Here, we simulate the potential observable compact object binaries in the vicinity of a SMBH under the effect of EKL via the GW interferometers. 

Various merger channels within the GC, besides the EKL assisted merger channel, also contributes to the formation of compact object binaries. Those include binary-single and binary-binary mediated interactions \citep[e.g.,][]{Rodriguez+18,Arcasedda+20} and single-single GW captures \citep[e.g.,][]{O'Leary+09,Tsang+13,Hoang+20}. Figure \ref{fig:snr} is agnostic to the merger channel of those compact object and highlights that a vast part of the parameter space can be detectable via LISA. 

However, we note that for NS-BH as well as BH-BH binaries, the EKL mechanism may be a significant contributor \citep[e.g.,][]{Hoang+18,Hoang+20}. Focusing on this merger process, we adopt the initial distribution from \citet{Stephan+19} post EKL distribution and generate a large population at the {\it onset of Roche limit crossing}. Using {\tt COSMIC}, we then evolve these binaries to their compact object stage. As expected, the compact object binary population near a SMBH has different properties, as highlighted in Figure \ref{fig:inihis}, than field binaries. In particular, these binaries possess shorter semi-major axis compared to isolated field binaries, due to interaction with passing objects at this dense environment \citep{Rose+20}. Additionally, we expect the binaries' eccentricity distribution to be excited due to the EKL mechanism \citep[e.g.,][]{Naoz+14}.

We have tested both solar and sub-solar metallicities. As expected the sub-solar metallicity produces more BH-BH binaries, by a factor $\sim 18$. The population of NS-NS binaries for solar metallicity is about doubled that of sub-solar metallicity. We note that although recent studies reported that the GC may have a super-solar population \citep[e.g.,][]{Feldemier-Krause-17,Do+18}, the hierarchical formation of galaxy \citep[e.g.,][]{White+78} suggested that sub-solar metallicities should exist at high abundance. Adopting sub-solar and solar metallicity allow us to also extrapolate the conditions of these sources to other galactic nucleus. The exploration of super-solar metallicity is beyond the scope of this paper.

From Figure \ref{fig:BH} and \ref{fig:NS}, we see that over the observation interval of 4 years, BH-BH binaries are the dominant observable sources with the highest SNR of GW signals (SNR $> 100$, Figure \ref{fig:BH}) while WD-WD binaries produce the most abundant amount of systems mostly clustered around SNR $\lesssim 10$ (Bottom row of Figure \ref{fig:NS}). Those two types of compact object provide the most promising GW sources of LISA via the EKL merging channel. Particularly, the WD-WD binaries, with their low separation and specific angular momentum (Figure \ref{fig:unihis}), will lead to a strong overlap of orbital frequency and the LISA sensitivity curve. Not only will they contribute to noise confusion but might also be detected individually. In comparison, NS-NS and NS-BH binaries are less likely to be observed within this formation channel but possibly abundant through other means, which we leave to explore in future study. 

 We also estimate the number of binaries, per galaxy, that are expected to be visible within the LISA band. While the details are highly uncertain, we are motivated by our own galaxy, for example, which has about a million stars within the inner parsec. We note that this is a conservative estimation because further from the SMBH, unbending of the binary due to interactions with neighboring stars is less efficient \citep{Rose+20}, thus we expect an even larger population than these conservative numbers. Overall we find, depending on the metallicity, about $140-150$ WD-WD, $0.2-2$ NS-BH, $2-4$ NS-NS, and $2-20$ BH-BH to be visible within the LISA band.

While we adopted a $D_l \sim 8$kpc for our GC, since the amplitude of GW signal is linear with the luminosity distance (Equation (\ref{eq:snr})), a detection of compact binaries in Milky Way like galaxies with distance $> 1 $~Mpc is possible with a longer observation timescale ($T_{\rm obs} \geq 4 {\rm yr}$). Hence, our results can also be extended to binary sources marginally observable via LISA at this luminosity distance. Particularly, \citet{Hoang+19} has shown that those systems have a chance to be detected via the eccentricity oscillations through the gravitational perturbation from the SMBH in the Galactic Nuclei, i.e.,  EKL. For a galaxy that contains a more massive SMBH, some systems might be visible via the eccentricity oscillation with only $T_{\rm obs} \sim 1$ yr \citep[e.g.,][]{Emami+20}.

The compact object binary population we found may also merge either via GW emission or EKL as time goes by. We roughly estimate the rate that the LIGO/Virgo can detect (see Table \ref{ligo}). These rates are sensitive to the SFR, where we assume a very conservative rate. However, E$+$A galaxies, or star burst galaxies, may undergo star formation episode that could possibly increase the stellar population by $\sim 10\%$, as well as increase the star tidal disruption events \citep[e.g.,][]{Dressler+83,Swinbank+12,Arcavi+14,Stone+16}.

We have shown that the GC, as well as other galactic nucleus, are potentially significant sources for LISA and LIGO/Virgo.  Most importantly, compact object binary at the galactic center can have an extremely large SNR in the LISA band.  Furthermore, the GW signal from the GC predicted here will have a preferential direction, compared to other detections in LISA, which may assist in disentangling the various signals.






\acknowledgments

We thank Salvo Vitale, Mark Morris and Tuan Do for useful discussion. 
HW thanks the UCLA-summer REU program. APS acknowledges partial support by the Thomas Jefferson Chair Endowment for Discovery And Space Exploration and partial support through the Ohio Eminent Scholar Endowment. HW, APS, SN and BMH acknowledge the partial support of NASA grants No. 80NSSC20K0505,  80NSSC19K0321 and NSF through grant No. AST- 1739160. SN thanks Howard and Astrid Preston for their generous support. 
%






\bibliography{main}{}
\bibliographystyle{aasjournal}



\end{document}